\documentclass[runningheads]{llncs}

\usepackage{graphicx}
\usepackage[colorlinks=true, allcolors=blue]{hyperref}

\usepackage{fixmath}
\usepackage{bm}
\usepackage{amsbsy}
\usepackage{color}
\usepackage{verbatim}
\usepackage{multirow}
\usepackage{amssymb}

\usepackage{array}
\usepackage{mathtools}
\usepackage{amsmath}
\usepackage{graphicx}
\usepackage{tikz}
\usetikzlibrary{positioning}
\usepackage{xcolor}

\usepackage{thm-restate}

\let\varepsilon=\varepsilon

\usepackage{booktabs}
\usepackage{changes}
\usepackage{amsmath, amsfonts, mathrsfs, mathtools}
\usepackage{changes}
\usepackage{enumitem}
\usepackage[linesnumbered,ruled,vlined,noend]{algorithm2e}
\usepackage{color}
\usepackage{bm}
\usepackage{graphicx}
\usepackage{layouts}
\newcommand{\mycite}[1]{\cite{#1}}

\newcommand{\myFullGame}{Facility Assignment Game with Fair Cost Sharing}
\newcommand{\myGame}{FAG-FCS}
\newcommand{\myGames}{FAG-FCS}

\begin{document}
\title{Facility Assignment with Fair Cost Sharing: Equilibrium and Mechanism Design}
\titlerunning{Facility Assignment with Fair Cost Sharing}
% If the paper title is too long for the running head, you can set
% an abbreviated paper title here
%
\author{Mengfan Ma, Mingyu Xiao, Tian Bai, Xin Cheng}
\authorrunning{M. Ma, M. Xiao, T. Bai, X. Cheng}
% First names are abbreviated in the running head.
% If there are more than two authors, 'et al.' is used.
%
\institute{University of Electronic Science and Technology of China, Chengdu, China \email{Mengfanma1@gmail.com, myxiao@uestc.edu.cn, tian.bai.cs@outlook.com, xincheng666888@gmail.com,}}
\maketitle              % typeset the header of the contribution
\begin{abstract}
    In the one-dimensional facility assignment problem, \(m\) facilities and \(n\) agents are positioned along the real line.
    Each agent will be assigned to a single facility to receive service.
    Each facility incurs a building cost, which is shared equally among the agents utilizing it.
    Additionally, each agent independently bears a connection cost to access a facility.
    Thus, an agent's cost is the sum of the connection cost and her portion of the building cost.
    The social cost is the total cost of all agents.
    Notably, the optimal assignment that minimizes the social cost can be found in polynomial time.
    In this paper, we study the problem from two game-theoretical settings regarding the strategy space of agents and the rule the assignment.
    In both settings, agents act strategically to minimize their individual costs.
    
    In our first setting, the strategy space of agents is the set of facilities, granting agents the freedom to select any facility. Consequently, the self-formed assignment can exhibit instability, as agents may deviate to other facilities.
    We focus on the computation of an \emph{equilibrium assignment}, where no agent has an incentive to unilaterally change her choice.
    We show that we can compute a pure Nash equilibrium in polynomial time.
    % Furthermore, we prove that the social cost of the computed equilibrium is a \(\ln n\)-approximation to that of the optimal assignment.

    In our second setting, agents must report their positions to a mechanism for assignment to facilities.
    Here, the strategy space of agents becomes the set of all positions.
    Our interest lies in \emph{strategyproof mechanisms}, which ensure no agent would misreport her position.
    It is essential to note that the preference induced by the agents' cost function is more complex as it depends on how other agents are assigned.
    We establish a strong lower bound against all strategyproof and anonymous mechanisms: none can achieve a bounded social cost approximation ratio.
    Nonetheless, we identify a class of non-trivial strategyproof mechanisms for any \(n\) and \(m\) that is unanimous and anonymous.
    %\keywords{Facility Assignment Game, Fair Cost Sharing, Equilibrium, Mechanism Design}
\end{abstract}

\section{Introduction}
In this paper, we investigate the one-dimensional \myFullGame{} (\myGame) from a game-theoretical perspective, exploring two settings with a focus on equilibrium computation and mechanism design, respectively. Generally, there are $n$ strategic agents and $m$ facilities (resources) positioned along the real line. Each agent will make a selection or be assigned to one facility to receive service. Each facility incurs a building cost, determined by its location, which is evenly distributed among the agents utilizing it. Additionally, each agent bears an individual connection cost to access the facility she uses. Consequently, an agent's cost is the sum of the connection cost and her proportionate share of the building cost. 
Each agent acts strategically within her strategy space and tries to minimize her own cost in the output assignment. 

The problem models many real-world scenarios, where agents jointly use resources that have load-dependent monetary costs and player-specific connection costs. The monetary cost of each resource must be shared fairly by its users while the player-specific connection cost is unavoidable physical quantities, such as the distance. In the one-dimensional context, this game can depict scenarios like the allocation of public infrastructure resources along a highway or the distribution of computing resources among entities in a computer bus.
Furthermore, there are various non-geographical settings that align with the problem, such as Content Distribution Network design, and distributed selfish caching \cite{chun2004selfish}. Additionally, the one-dimensional game may also find applications in social choice, where all possible assignments represent candidates, and the positions of the agents reflect their preferences over these assignments.

It is well-known that the optimization problem seeking an assignment to minimize the social cost can be solved in polynomial time \cite{love1976note}. However, it's worth noting that such an optimal assignment may not necessarily align with the individual interests of the agents. To address this issue, we explore the problem in two strategic settings.

In our first setting, no centralized authority can enforce an assignment and agents can select any facility on their own, i.e., the strategy space of each agent is the set of facilities. 
This model corresponds to the one-dimensional version of the facility location game studied by \cite{hansen2008pure,hansen2009improved}.
In this context, the concept of pure Nash equilibrium (PNE) holds particular significance, representing a stable state in which no agent has an incentive to unilaterally change her strategy.
However, in more general cost-sharing games, computing a PNE is a challenging task, possibly even PLS-complete \cite{syrgkanis2010complexity}. 
Prior research in \cite{hansen2008pure,hansen2009improved} has provided polynomial-time algorithms to approximate PNEs, but the computational complexity of finding an exact PNE in metric \myGames{} remains an open question. 
Our focus lies in understanding the computation of equilibria in the one-dimensional setting and assessing the equilibrium's quality relative to the global optimum.

In the second setting, we assume there is an authority capable of determining assignments, while the precise positions of agents remain \textit{private}, known exclusively to the agents themselves. The authority can only access the positions reported by the agents.
A mechanism or algorithm is strategyproof if agents perceive it to be in their best interest to accurately report their positions to the mechanism.
However, a strategyproof mechanism may not necessarily achieve the optimal social cost.
To quantify the inherent trade-off between strategyproofness and optimality, we turn to the notion of approximation ratio \cite{procaccia2013approximate}.
In almost all previous work on truthful assignment or facility location (see, e.g., \cite{lu2009tighter,ma2023facility,anshelevich2016blind}), the cost of an agent is only determined by her assigned facility or matched endpoint.
Our goal is to understand this trade-off and explore the limits of approximation ratio achievable strategyproof mechanisms and design mechanisms of desirable axiomatic properties.

\subsection{Contributions}
In the first setting, where the strategy space of each participant is the set of facilities, we devise a dynamic programming algorithm that computes a PNE in polynomial time for any \(n\) and \(m\), leveraging the decomposability of structure in PNEs and the potential function. Remarkably, this marks the first polynomial-time algorithm for exact PNEs in \myGame{}, despite the one-dimensional constraint. Additionally, we demonstrate that the PNE attained through this algorithm approximates the optimal solution with a factor of \(\ln n\).

In a setting where strategic agents report their positions to a mechanism for facility assignment, we provide a complete characterization of strategyproof and anonymous mechanisms for the scenario where \(m=n=2\). Despite the limited scope of \(m=n=2\), we establish a strong lower bound, demonstrating that no strategyproof and anonymous mechanism can guarantee a bounded approximation ratio even with fixed \(m\) and \(n\). This revelation suggests a theoretical paradox: in even the simplest congestion games, strategyproofness inherently conflicts with social cost optimization.
Inspired by the characterization, we design a class of non-trivial strategyproof mechanisms for any \(n\) and \(m\), which are both unanimous and anonymous. This implies that the preference domain induced by our cost function escapes the Gibbard-Satterthwaite impossibility result.
We believe that our characterization and mechanisms will offer valuable insights into this novel preference domain.

\subsection{Related Work}
\paragraph{Complexity of PNEs in Congestion Games}  
The existence of a PNE has been confirmed by Rosenthal \mycite{rosenthal1973class}, showing that all improvement sequences possess finite lengths. 
Fabrikant et al. \mycite{fabrikant2004complexity} show, however, that these improvement sequences may require an exponential number of iterations. 
Their analysis reveals a connection between congestion games and local search problems, establishing that it is PLS-complete to compute a PNE when the strategies of players are paths in a network (referred as to Network Congestion Game) with non-decreasing cost functions. 
For network congestion games with fair cost sharing, in which the cost function is obviously non-increasing, Syrgkanis \mycite{syrgkanis2010complexity} demonstrates that it is PLS-complete to compute a PNE.

On the positive side, Fabrikant et al. \mycite{fabrikant2004complexity} show that in Network Congestion Game with a non-decreasing cost function, a PNE can be found in polynomial time when all players have the same source and sink.
Furthermore, Ackermann el al. \mycite{ackermann2008impact} introduce the Matroid Congestion Game, where the strategy space of each player consists of the bases of a matroid over the set of resources. They prove that any best response dynamics in this setting converge in polynomially many steps.

As previously mentioned,  the game studied in the first setting can be viewed as the one-dimensional version of the Metric \myGame{} in Hansen and Telelis \cite{hansen2008pure,hansen2009improved}.
They demonstrate that the best response dynamics may require super-polynomial steps to converge and provide polynomial-time algorithms to compute approximate PNEs.
To the best of our knowledge, there are no hardness results or polynomial-time algorithms available for computing an exact PNE in the metric \myGame{}. 
Therefore, our algorithmic findings can be considered the initial step towards understanding the complexity of PNEs in the metric \myGame{}.

Moreover, Milchtaich \cite{milchtaich1996congestion} examines congestion games with player-specific cost functions and demonstrates the existence of at least one PNE in such games. Our model can also be construed as a particularization of Milchtaich's framework, where the individual distance serves as the player-specific cost function. 

\paragraph{Mechanism Design without Money} Our work in the mechanism design setting belongs to the realm of \emph{approximate mechanism design without money}, a term coined by Procaccia and Tennenholtz \mycite{procaccia2013approximate} to describe problems where a certain objective function is optimized under the hard constraints imposed by the requirement of strategyproofness.
In their setting, facilities' locations are not predetermined but instead determined by a mechanism, and an agent's cost is the distance to her nearest facility.
This work has attracted numerous follow-up research, on the same model \cite{lu2009tighter,lu2010asymptotically,alon2010strategyproof,fotakis2014power}, and on its variants \cite{aziz2020facility,li2021budgeted,ma2023facility}.
It is worth noting that Ma et al. \cite{ma2023facility} introduce a model in which facilities charge their customers an entrance fee, conceptually resembling the share of building cost in our model. However, a critical distinction lies in their model, where the entrance fee of a facility remains fixed, resulting in an agent's cost being solely contingent on the location of her chosen facility. It is notable that prevailing literature focuses on the one-dimensional setting, as evidenced in \cite{procaccia2013approximate,fotakis2014power,lu2010asymptotically,li2021budgeted,ma2023facility}.

Another closely related line of research pertains to strategyproof assignment or matching mechanisms without money. 
Caragiannis et al. \mycite{caragiannis2022truthful} focus on resource augmentation within the strategyproof facility assignment problem, given facility capacity limits.
Dughmi and Ghosh \mycite{dughmi2010truthful} investigate strategyproof mechanism design in the generalized assignment problem, where agents' private information becomes the existence of an edge.
Anshelevich and Sekar \mycite{anshelevich2016blind} study strategyproof matching and clustering mechanisms in metric spaces, constrained by the mechanism's access solely to ordinal preference information.

It is crucial to note that, in all the aforementioned literature, an agent's cost is solely determined by her assigned facility or matched endpoint. Consequently, our mechanism design setting fundamentally deviates from prior research.

\section{Preliminaries}
We introduce several notations for \myGame{}, which is commonly employed in both settings.
Denote by \([n]\) the set \(\{1,2,\ldots, n\}\) for any \(n\in \mathbb{N}\). We have \(n\) \emph{agents} on the real line with position profile \(\mathbf{x}=(x_1,x_2 \ldots, x_n)\) where \(x_i\in \mathbb{R}\) for \(i\in [n]\), and \(m\) \emph{facilities} on the real line with position profile \(\bm{\ell}=(\ell_1,\ell_2, \ldots, \ell_m)\).
The \emph{building cost} of facility \(j\in [m]\) is \(b_j\in \mathbb{R}_{>0}\). Denote by \(\mathbf{b}=(b_1,b_2,\ldots, b_m)\) the building cost profile.
We call the configuration of facilities \(\mathbf{e}=(\textbf{b},\bm{\ell})\) \emph{environment}. Let \(\mathcal{E}\) be the set of all possible environments.
An \emph{assignment} \(\mathbf{s}=(s_1,s_2,\ldots,s_n)\in [m]^n\) specifies the facility assigned to each agent.
Given an assignment \(\mathbf{s}\), for \(j\in [m]\), define \(n_j(\mathbf{s})\coloneqq |\{i\in [n]|s_i=j\}|\), that is, \(n_j(\mathbf{s})\) is the number of agents assigned to facility \(j\in [m]\). 
Given an environment \(\mathbf{e}\), the \emph{cost} of agent \(i\) is 
\begin{align*}
    c(x_i,\mathbf{s},\mathbf{e})\coloneqq d(x_i,s_i)+\frac{b_{s_i}}{n_{s_i}(\mathbf{s})},
\end{align*}
where \(d(x_i,s_i)=|x_i-\ell_{s_i}|\).
Then the \emph{social cost} of assignment \(\mathbf{s}\) is \(\mathrm{SC}(\mathbf{x},\mathbf{s},\mathbf{e})\coloneqq \sum_{i=1}^{n} c(x_i,\mathbf{s},\mathbf{e})\).
For a profile \(\mathbf{x}\), let \(OPT(\mathbf{x},\mathbf{e})\coloneqq \min_{\mathbf{s}\in [m]^n}\mathrm{SC}(\mathbf{x},\mathbf{s},\mathbf{e})\) be the optimal (smallest) social cost.
With a slight abuse of notation, when the environment \(\mathbf{e}\) is evident from context, we may also denote \(c(x_i,\mathbf{s},\mathbf{e}), \mathrm{SC}(\mathbf{x},\mathbf{s},\mathbf{e})\) and \(OPT(\mathbf{x},\mathbf{e})\) as \(c(x_i,\mathbf{s}), \mathrm{SC}(\mathbf{s},\mathbf{e})\) (or \(\mathrm{SC}(\mathbf{s})\)) and \(OPT(\mathbf{x})\), respectively.

Given an environment $\mathbf{e}$, an agent position profile $\mathbf{x}$ and an assignment $\mathbf{s}$, we use the ratio $\mathrm{SC}(\mathbf{s},\mathbf{e})/OPT(\mathbf{x},\mathbf{e})$ to evaluate the quality of an assignment with respective to $\mathbf{x}$.

The proofs for statements marked with \(\clubsuit\) are not included in the main text but can be found in the appendix, which spans 13 pages.

\section{The Computation of PNEs}
In this section, we consider our first setting where the strategy space of each agent \(i\) is \([m]\). That is, \(i\) chooses a facility \(j\in [m]\) to receive service from.
An assignment \(\mathbf{s}\) is also called a \emph{strategy profile} in this setting. 
\begin{definition} [pure Nash equilibrium] \label{def_pne}
    A strategy profile $\mathbf{s}$ is a PNE if no player could decrease her cost by unilaterally changing her chosen facility.
    Formally, \(\mathbf{s}\) is a PNE if for any $i\in [n]$ and \(s_i'\in [m]\), we have \(c(x_i,\mathbf{s})\le c(x_i,\mathbf{s}')\), where $\mathbf{s}'=(s_i',\mathbf{s}_{-i})$ is obtained from $\mathbf{s}$ by replacing
    the strategy of \(i\) in $\mathbf{s}$ with $\mathbf{s}_i'$.
\end{definition}

The main result in this section is the following theorem.

\begin{theorem} \label{thm_pne_poly}
    A PNE of \myGames{} can be computed in \(\mathcal{O}(n^2m^2)\).
\end{theorem}

\subsection{Congestion Game and Potential Function} \label{apd_cg}
% Congestion games are an important class of games in game theory. They represent situations that commonly occur in roads, communication networks, oligopoly markets, and natural habitats. There is a set of resources and several players who need resources; each player chooses a subset of these resources; the cost in each resource is determined by the number of players whose strategies contain this resource. The cost of each player is the sum of costs among all resources he chooses. Naturally, each player wants to minimize his own cost; however, each player's choice imposes a negative or positive externality on the other players, which may lead to inefficient outcomes. We give the formal definition as follows:

Congestion games are an important class of games in game theory. There is a set of resources and several players who need resources; each player chooses a subset of these resources; the cost in each resource is determined by the number of players whose strategies contain this resource.We give the formal definition as follows:

\begin{definition} [congestion game]
    A Congestion Game
    % denoted by  $$\left\langle N, R,\left(S_i\right)_{i \in N},\left(f_r\right)_{r \in R}\right\rangle,$$  
    consists of a set of players $N$, a set of resources $R$. For each player \(i\) there is a set of strategies $S_i \subseteq 2^R$. For each resource $r\in R$ there is a cost function $f_r: \mathbb{N}\rightarrow \mathbb{R}_{\ge 0}$, which maps the number of players using resource \(r\) to a non-negative cost.
    The cost of a player $i$ at a strategy profile $\mathbf{s}$ is $c_i(\mathbf{s}):=\sum_{r \in s_i} f_r(n_r(\mathbf{s})),$ where $n_r(\mathbf{s})$ is the number of players using resource $r$ in strategy profile \(\mathbf{s}\).
    Thus, for a given strategy profile $\mathbf{s}$, the Social Cost of $\mathbf{s}$ is:
    \begin{align*}
        SC(\mathbf{s}) =\sum_{i \in N} c_i(\mathbf{s}) = \sum_{i \in N} \sum_{r \in s_i} f_r(n_r(\mathbf{s}))
        = \sum_{r \in R} \left(f_r(n_r(\mathbf{s}))\cdot n_r(\mathbf{s})\right).
    \end{align*}
\end{definition}

Rosenthal \cite{rosenthal1973class} proves that every congestion game has a PNE.
The key is to define the \emph{potential function} for each congestion game:
$\Phi(\mathbf{s})=\sum_{r \in R} \sum_{k=1}^{n_j(\mathbf{s})} f_r(k),$ where \(\mathbf{s}\) is a strategy profile.
Note that this function is not the social cost, but rather a discrete integral of sorts.
Then he showed that the potential function has the critical property that, for every unilateral deviation by an arbitrary agent, the change in the potential function values equals the change in the deviator's cost. Formally, for every strategy profile \(\mathbf{s}\), and unilateral deviation \(s_i'\in S_i\), it holds that \(\Phi(s_i',\mathbf{s}_{-i}) - \Phi(\mathbf{s}) = c_i(s_i',\mathbf{s}_{-i}) - c_i(\mathbf{s})\).
% \begin{align*}
%     \Phi(s_i',\mathbf{s}_{-i}) - \Phi(\mathbf{s}) = c_i(s_i',\mathbf{s}_{-i}) - c_i(\mathbf{s})
% \end{align*}

Now observe that any strategy profile that minimizes \(\Phi\) is a PNE: Fixing all but one player, any improvement in strategy by that player corresponds to decreasing 
\(\Phi\), which cannot happen at a minimum. Now since there is a finite number of possible strategy profiles and each is monotone, there exists a PNE. The existence of a potential function has an additional implication, called the \emph{finite improvement property}: If we start with any strategy-vector, pick a player arbitrarily, and let him change his strategy to a better strategy for him, and repeat, then the sequence of improvements must be finite (that is, the sequence will not cycle). This is because each such improvement strictly decreases the potential function.
This potential function is the main tool until today to show the existence of PNEs and to discuss the quality of PNEs.

\begin{theorem}(\cite{rosenthal1973class})
    Every congestion game possesses at least one PNE.
\end{theorem}

Our \myGames{} specializes in the congestion game since we can consider both facilities and the edge between facilities and agents as resources. 
Note that there may exist a facility that no agent selects. We denote \(F_{\mathbf{s}}\) the set of facilities that is chosen by at least one agent.
Thus, according to \mycite{rosenthal1973class}, given a strategy profile \(\mathbf{s}\), the potential function in one-dimensional \myGame{} is 
\begin{align} \label{eq_pfunc1}
    \Phi(\mathbf{s})= \sum_{j\in F_\mathbf{s}}\sum_{k=1}^{n_j(\mathbf{s})}\frac{b_j}{k} + \sum_{i=1}^{n} |x_i-\ell_{s_i}|.
\end{align}
The potential function has the critical property that, for every unilateral deviation by some agent, the change in the potential function values equals the change in the deviator's cost. Formally, for every strategy profile \(\mathbf{s}\), and unilateral deviation \(s_i'\in [m]\), it holds that \(\Phi(\mathbf{s})-\Phi(\mathbf{s'}) = c(x_i,\mathbf{s})-c(x_i,\mathbf{s}'),\)
% \begin{align*}
%     \Phi(\mathbf{s})-\Phi(\mathbf{s'}) = c(x_i,\mathbf{s})-c(x_i,\mathbf{s}'), 
% \end{align*}
where \(\mathbf{s}'=(s_i,\mathbf{s}_{-i})\). It is easy to see that a strategy profile minimizing the potential function must be a PNE. 
Let \(A_j(\mathbf{s})\) be the set of agents who select facility \(j\in F_{\mathbf{s}}\).
Then we can derive a more compact form of the potential function from \eqref{eq_pfunc1}:
\begin{align} \label{eq_pfunc2}
    \Phi(\mathbf{s}) = \sum_{j\in F_\mathbf{s}}\left(\sum_{k=1}^{n_j(\mathbf{s})}\frac{b_j}{k}+\sum_{i\in A_j(\mathbf{s})}|x_i-\ell_j|\right).
\end{align}

Next, we show how to find in polynomial time a strategy profile that minimizes \eqref{eq_pfunc2}. 
We have the following ``no-cross'' property of any PNE in one-dimensional \myGames{}.
\begin{lemma}[\(\clubsuit\)] \label{lm_pne_nocross}
    For any PNE \(\mathbf{s}\) and any two agents \(a,b\in [n]\),  we have \(\ell_{s_a}\le\ell_{s_b}\) if \(x_a < x_b\).
\end{lemma}

We say \(S\subseteq [n]\) is \emph{consecutive} if \(S=\{x_{i_1},x_{i_1+1}, \ldots, x_{i_2}\}\) for some \(1\le i_1 \le i_2 \le n\).
By Lemma \ref{lm_pne_nocross} we immediately have the following lemma that depicts the consecutive structure of any PNE.
\begin{lemma} \label{lm_pne_consecutive}
    For any PNE \(\mathbf{s}\), let \(a,b,c\in [n]\) be three agents such that \(x_a<x_b<x_c\). Then we have \(s_b=s_a=s_c\) if \(s_a=s_c\).
\end{lemma}

Equipped with lemmas \ref{lm_pne_nocross}, \ref{lm_pne_consecutive} and the potential function \eqref{eq_pfunc2} we can prove Theorem \ref{thm_pne_poly}. The proof is omitted here and can be found in Appendix \ref{app_pne_poly}.

Next we show that the social cost of the PNE obtained by Algorithm \ref{alg_pneDp} is at most \(\ln n\) times that of the optimal assignment.
\begin{proposition}[\(\clubsuit\)]\label{prop_pne_lnn}
    For any instance of the one-dimensional \myGame{}, let \(\hat{\mathbf{s}}\) be the PNE obtained by Algorithm \ref{alg_pneDp} and \(\mathbf{s}^*\) be the assignment that minimizes the social cost. We have \(\mathrm{SC}(\hat{\mathbf{s}}) \le \ln n \cdot \mathrm{SC}(\mathbf{s}^*)\).
\end{proposition}
% \begin{proof}
%     We can rewrite the social cost of \(\hat{\mathbf{s}}\) as 
%     \begin{align*}
%         \mathrm{SC}(\hat{\mathbf{s}})=\sum_{j\in F(\hat{\mathbf{s}})} b_j + \sum_{i=1}^{n} d(x_i,\hat{s}_i)
%     \end{align*}
%     Compare above equation with the potential function \eqref{eq_pfunc1}, we get \(\mathrm{SC}(\hat{\mathbf{s}}) \le \Phi(\hat{\mathbf{s}})\). 
%     Let \(\mathcal{H}_k=(1+1/2+\cdots+1/k)\) be the \(k\)th harmonic number, by \eqref{eq_pfunc1} we have 
%     \begin{align*}
%         \Phi(\mathbf{\mathbf{s}^*})&= \sum_{j\in F(\mathbf{\mathbf{s}^*})}\sum_{k=1}^{n_j(\mathbf{s}^*)}\frac{b_j}{k} + \sum_{i=1}^{n} d(x_i,s^*_i)\\
%         & \le \mathcal{H}_n \cdot\sum_{j\in F(\mathbf{s}^*)} b_j + \sum_{i=1}^{n} d(x_i,s^*_i)\\
%         & \le \mathcal{H}_n \cdot(\sum_{j\in F(\mathbf{s}^*)} b_j + \sum_{i=1}^{n} d(x_i,s^*_i)) \\
%         & \le \ln n \cdot \mathrm{SC}(\mathbf{s}^*).
%     \end{align*}
%     By definition of \(\hat{\mathbf{s}}\) we also have \(\Phi(\hat{\mathbf{s}})\le \Phi(\mathbf{s}^*)\), which leads to 
%     \begin{align*}
%         \mathrm{SC}(\hat{\mathbf{s}}) \le \Phi(\hat{\mathbf{s}}) \le \Phi(\mathbf{s}^*) \le \ln n\cdot \mathrm{SC}(\mathbf{s}^*).
%     \end{align*}

% \hfill$\square$
%\end{proof}

\section{Characterization and Design of Strategyproof Mechanisms}

In our mechanism design setting, each agent's position is her private information. 
A mechanism outputs an assignment of agents to facilities according to a reported location profile \(\mathbf{x}\) and a given environment \(\mathbf{e}\), and thus is a function \(f:\mathbb{R}^n\times \mathcal{E}\rightarrow [m]^n\). 
With a slight abuse of notation, when the environment \(\mathbf{e}\) is explicitly provided, we may also write \(f(\mathbf{x},\mathbf{e})\) as \(f(\mathbf{x})\).
Given that an agent might misreport her location to decrease her cost, it is necessary to design \emph{strategyproof} mechanisms.
\begin{definition} [strategyproofness]
    A mechanism \(f\) is strategyproof if no agent can benefit from misreporting her position. Formally, given any \(\mathbf{e}\in \mathcal{E}\), any agent \(i\in N\), profile \(\mathbf{x}=(x_i,\mathbf{x}_{-i})\in \mathbb{R}^n\), and misreported location \(x_i'\in \mathbb{R}\), it holds that
    \begin{align*}
        c(x_i,f(\mathbf{x},\mathbf{e})) \le c(x_i, f((x_i',\mathbf{x}_{-i}),\mathbf{e})).
    \end{align*}
\end{definition}

Next, we introduce \emph{anonymity}, which is a fundamental fairness property that requires all agents to be treated alike.
\begin{definition} [anonymity]
    A mechanism \(f\) is anonymous if for any environment \(\mathbf{e}\), any \(\mathbf{x}\in \mathbb{R}^n\) and any permutation \(\pi\) over $[n]$, we have \(f(\mathbf{x},\mathbf{e})=f((x_{\pi(1)},x_{\pi(2)},\ldots, x_{\pi(n)}),\mathbf{e})\).
\end{definition}
By the above definition, in an anonymous mechanism, agents with exactly the same positions must be assigned to the same facility. We use \emph{approximation ratio} to evaluate the performance of a mechanism.
The approximation ratio of a mechanism is the maximum ratio $\frac{\mathrm{SC(f(\mathbf{x},\mathbf{e}), \mathbf{e})}}{OPT(\mathbf{x},\mathbf{e})}$ over all $\mathbf{e}\in \mathcal{E}$ and $\mathbf{x}\in \mathbf{R}^n$.

The main result in this section is the following theorem.

\begin{theorem} \label{thm: lb}
    No strategyproof and anonymous mechanism can achieve a bounded approximation ratio for the social cost, even when restricted to $m = n = 2$.
\end{theorem}

We prove Theorem~\ref{thm: lb} by giving a complete characterization of mechanisms for $m = n = 2$ in the next subsection. 

\subsection{Characterization of Strategyproof and Anonymous Mechanisms for \(m=n=2\)} \label{subsec: char}

For \(m=2\), let $\Delta \coloneqq \ell_{2} - \ell_{1}$ be the distance between two facilities.
First, we derive several useful properties of any strategyproof and anonymous mechanism.
Denote by \(f_i(\mathbf{x},\mathbf{e})\) the facility assigned to agent \(i\) by mechanism \(f\) with environment \(\mathbf{e}\) and location profile \(\mathbf{x}\).

\begin{lemma}[\(\clubsuit\)] \label{lm_sp_ppt}
    Let $f$ be a strategyproof and anonymous mechanism. For any given environment \(\mathbf{e}\), and any two agent location profiles \(\mathbf{x}=(x_1,x_2,\ldots, x_n)\), \(\mathbf{x}'=(x_i', \mathbf{x}_{-i})\), let \(j=f_i(\mathbf{x},\mathbf{e})\) and \(j'=f_i(\mathbf{x}',\mathbf{e})\). Let \(n_j=n_j(f(\mathbf{x},\mathbf{e}))\) and \(n_{j'}'=n_{j'}(f(\mathbf{x}',\mathbf{e}))\).
    Then we have the following properties:

    %\textbf{\textup{(P1)}} if \(x_i<x_i'\) and \(I(x_i,x_i')\cap I(\ell_j,\ell_{j'})\ne \emptyset\), then \(\ell_j\le \ell_{j'}\);
    \textbf{\textup{(P1)}} {if \(x_i<x_i'\) and $\ell_{j'} < \ell_j$, then either $x'_{i} \leq \ell_{j'}$ or \(\ell_{j} \leq x_{i}\) holds;}

    \textbf{\textup{(P2)}} \(\displaystyle d(x_i',\ell_{j'})-d(x_i',\ell_j) \le \frac{b_j}{n_j}-\frac{b_{j'}}{{n'_{j'}}} \le d(x_i,\ell_{j'})-d(x_i,\ell_{j})\);

    \textbf{\textup{(P3)}} if \(j=j'\), then \(n_j=n_{j}'\).
\end{lemma}

The three properties in the above lemma hold for general \(m\) and \(n\).
(P2) and (P3) hold even for general spaces which can be non-metric.
When restricted in one-dimensional space, (P1) says that in some situations, if an agent moves towards or beyond her designated facility, then her designated facility will also move along the same direction.
Although in a very general form, (P2) is very useful in some cases.
For instance, when \(x_i<x_i'\), (P2) gives us \({b_j}/{n_j}-{b_{j'}}/{n_{j'}'}=\Delta\).
This observation is used extensively in our characterization.

Next, we focus on the case where \(m=n=2\), i.e., the case where the mechanism needs to assign two agents to two facilities.
We will see that the characterization of strategyproof and anonymous mechanisms for this simple case is highly non-trivial.
We first derive some useful lemmas for this scenario.

\begin{lemma}[\(\clubsuit\)] \label{lm_sp_ppt_2a}
  Let \(f\) be a strategyproof and anonymous mechanism.
  For a given environment \(\mathbf{e}=((b_1, b_2),(\ell_1, \ell_2))\) and an \(2\)-agent location profile \(\mathbf{x} = (x_{1}, x_{2})\) with \(x_1< x_2\),  we have

  \textbf{\textup{(P4)}} if intervals \((x_1,x_2)\cap (\ell_1,\ell_{2})= \emptyset\), then \(f_1(\mathbf{x})=f_2(\mathbf{x})\);

  \textbf{\textup{(P5)}} if intervals \((x_1,x_2)\cap (\ell_1,\ell_{2})\ne \emptyset\), then \(f(\mathbf{x})\ne (2,1)\).
\end{lemma}

\noindent
In order to characterize the strategyproof and anonymous mechanisms, we need to define the following three positions related to the environmental parameters.
\begin{definition}
    Given an environment \(\mathbf{e}=((b_1, b_2),(\ell_1, \ell_2))\), we define three environmental parameters:
    \begin{equation*}
        \left\{
        \begin{aligned}
        L(\mathbf{e}) := \frac{1}{2}\Delta + \frac{1}{4}b_{2} - \frac{1}{2}b_{1}; \\
        M(\mathbf{e}) := \frac{1}{2}\Delta + \frac{1}{4}b_{2} - \frac{1}{4}b_{1}; \\
        R(\mathbf{e}) := \frac{1}{2}\Delta + \frac{1}{2}b_{2} - \frac{1}{4}b_{1}. \\
        \end{aligned}
        \right.
      \end{equation*}
\end{definition}
To simplify the notation, we also write the above three positions as $L$, $R$, and $M$ when the environment \(\mathbf{e}\) is specified.
It is easy to see that \(L<M<R\).
Next we introduce another position determined by \(f(x_1,x_2)\) with \(x_1=x_2\).
\begin{definition}
    Let $f$ be a strategyproof and anonymous mechanism.
    Given an environment \(\mathbf{e}=((b_1, b_2),(\ell_1, \ell_2))\), define \(x^{*} \coloneq \sup\{ x : f(x, x) = (1, 1) \}\).
    % \begin{equation*}
    %     x^{*} = \sup\{ x : f(x, x) = (1, 1) \}. 
    % \end{equation*}
\end{definition}
Note that \(x^*\in [-\infty,+\infty]\).
When there is no \(x\in \mathbb{R}\) such that \(f(x,x)=(1,1)\) we have \(\{x:f(x,x)=(1,1)\}=\emptyset\) and \(x^*=sup\: \emptyset = -\infty\). And when \(f(x,x)=(1,1)\) for any \(x\in \mathbb{R}\), we have \(x^*=\sup \mathbb{R}=+\infty\).

\begin{lemma}[\(\clubsuit\)] \label{lm_div_ppt}
    Let $f$ be a strategyproof and anonymous mechanism. 
    If there exists a profile \((\hat{x}_1,\hat{x}_2)\) with \(\hat{x}_1< \hat{x}_2\) such that \(f_1(\hat{x}_1,\hat{x}_2)\ne f_2(\hat{x}_1,\hat{x}_2)\), then
    \begin{align} \label{diag}
        \begin{cases}
            f_{2}(x, \hat{x}_{2}) = 2,~\forall x \le \hat{x}_2; \\
            f_{1}(\hat{x}_1, x) = 1,~\forall x \ge \hat{x}_1.
        \end{cases}
    \end{align}
\end{lemma}

\begin{lemma}[\(\clubsuit\)] \label{lm_mne0}
    Let $f$ be a strategyproof and anonymous mechanism.
    For any environment \(\mathbf{e}=((b_1, b_2),(\ell_1, \ell_2))\), if \(M\notin \{0,\Delta\}\), then 
    \begin{enumerate} [label=(\roman*)]
        \item there exists some \(j\in [2]\) such that \(f(x_1,x_2)=(j,j)\) for all \(x_1\le x_2 \le \ell_1\);
        \item there exists some \(j'\in [2]\) such that \(f(x_1,x_2)=(j',j')\) for all \(\ell_2\le x_1\le x_2 \).
    \end{enumerate}
    % if \(M\ne \Delta\), then 
    % \begin{enumerate} [label=(\roman*), start=3]
    %     \item there exists some \(j\in [2]\) such that \(f(x_1,x_2)=(j,j)\) for all \(x_1\le x_2 \le \ell_1\);
    %     \item there exists some \(j'\in [2]\) such that \(f(x_1,x_2)=(j',j')\) for all \(\ell_2\le x_1\le x_2 \).
    % \end{enumerate}
\end{lemma}

We define \(5\) types of mechanisms.
\begin{definition} \label{def_sp_mech}
    Let $f$ be a strategyproof and anonymous mechanism.
    \begin{enumerate}
        \item Type I: $f(\mathbf{x}) = (1, 1)$ for any $\mathbf{x}\in \mathbb{R}^2$ or $f(\mathbf{x}) = (2, 2)$ for any profile $\mathbf{x}\in \mathbb{R}^2$. This is the trivial strategyproof and anonymous mechanism that always assigns all agents to the same facility;
        \item Type II: $M = 0$ and 
            \begin{equation*}
                f(\mathbf{x}) = \left\{
                \begin{aligned}
                    &(1, 1) \text{~or~} (2, 2), &&\text{~if~} x_{1} = x_{2} \le \ell_{1}; \\
                    &f(x_{1}, x_{1}), &&\text{~if~} x_{1} \le \ell_{1}\text{~and~}x_{1} < x_{2}; \\
                    &(2, 2), &&\text{~if~}  x_{1} > \ell_{1}. \\
                \end{aligned}
                \right.
            \end{equation*}
        \item Type III: $M = \Delta$ and
            \begin{equation*}
                f(\mathbf{x}) = \left\{
                \begin{aligned}
                    &(1, 1), &&\text{~if~} x_{2} < \ell_{2}; \\
                    &(1, 1) \text{~or~} (2, 2), &&\text{~if~} \ell_{2} \leq x_{1} = x_{2}; \\
                    &f(x_{1}, x_{1}), &&\text{~if~} x_{2} \ge \ell_{2}\text{~and~}x_{1} < x_{2}; \\
                \end{aligned}
                \right.
            \end{equation*}
        \item Type IV: $0 < M < \Delta$ and
            \begin{equation*}
                f(\mathbf{x})=
                \left\{
                \begin{aligned}
                & (1, 1), && \text{~if~}x_{1} < M + \ell_{1}; \\
                & (1, 1)\text{~or~}(2, 2), && \text{~if~} x_{1} = x_{2} = M + \ell_{1}; \\
                & f(M + \ell_{1}, M + \ell_{1}), && \text{~if~} x_{1} = M + \ell_{1}; \\                    
                & (2, 2), && \text{~if~}x_{1} > M + \ell_{1}. \\
                \end{aligned}
                \right.
            \end{equation*}
        \item Type V: $0 < M < \Delta$ and
            \begin{equation*}
                f(\mathbf{x})=
                \left\{
                \begin{aligned}
                & (1, 1), && \text{~if~} x_{2} < M + \ell_{1}; \\
                & (1, 1)\text{~or~}(2, 2), && \text{~if~} x_{1} = x_{2} = M + \ell_{1}; \\
                & f(M + \ell_{1}, M + \ell_{1}), && \text{~if~} x_{2} = M + \ell_{1}; \\
                & (2, 2), && \text{~if~} x_{2} > M + \ell_{1}. \\
                \end{aligned}
                \right.
            \end{equation*}
    \end{enumerate}
\end{definition}

\begin{lemma}[\(\clubsuit\)] \label{lm_mech_sp}
    The mechanisms in Definition \ref{def_sp_mech} are strategyproof and anonymous.
\end{lemma}

\begin{theorem}[\(\clubsuit\)] \label{thm: char}
    Let $f: \mathbb{R}^2\rightarrow [2]^2$ be a mechanism for $m=n=2$. 
    \(f\) is strategyproof and anonymous if and only if \(f\) is one of the mechanisms in Definition \ref{def_sp_mech}.
    Given \(\mathbf{e}=((b_1, b_2),(\ell_1, \ell_2))\), the relation between the environmental parameters and the types of the mechanisms are demonstrated in Table \ref{tab_relation}.
    
    \begin{table}[t]
    \centering
    \caption{\small Relation between the environmental parameters and the types of the mechanisms. For example, it states that when \(2\Delta = b_1-b_2\), any strategyproof and anonymous mechanism must be type I or II.} \label{tab_relation}
    \begin{tabular}{cccccccc}
        \toprule
        Environmental parameters                  & I          & II         & III        & IV          & V         \\
        \midrule
        $2\Delta < b_{1} - b_{2}$                 & \checkmark &            &            &            &            \\
        $2\Delta = b_{1} - b_{2}$                 & \checkmark & \checkmark &            &            &            \\
        $b_{1} - b_{2} < 2\Delta < b_{2} - b_{1}$ & \checkmark &            &            & \checkmark & \checkmark \\
        $2\Delta = b_{2} - b_{1}$                 & \checkmark &            & \checkmark &            &            \\
        $2\Delta > b_{2} - b_{1}$                 & \checkmark &            &            &            &            \\
        \bottomrule
    \end{tabular}
    
    \end{table}
\end{theorem}

By the characterization in Theorem~\ref{thm: char}, we have the following corollary:

\begin{corollary}[\(\clubsuit\)] \label{cor_apx_n2}
    Let $f: \mathbb{R}^2\rightarrow [2]^2$ be any strategyproof and anonymous mechanism for $m=n=2$. 
    Given an environment \(\mathbf{e}=((b_1, b_2),(\ell_{1}, \ell_{2}))\) with \(0 < M < \Delta\), we have that the social cost approximation ratio of \(f\) is no less than
    \begin{align*}
        \max\left\{\frac{\min\{b_1,b_2\} + \Delta}{b_{1} + b_{2}}, \frac{\min\{b_1,b_2\} + \Delta + M}{\max\{b_{1}, b_{2}\} + M}\right\}
    \end{align*}
    % \begin{equation*}
    %     \left\{
    %     \begin{aligned}
    %         &\max\left\{\frac{\min\{b_1,b_2\} + \Delta}{b_{1} + b_{2}}, \frac{\min\{b_1,b_2\} + \Delta + M}{\max\{b_{1}, b_{2}\} + M}\right\}, &&\text{~if~} 0 < M < \Delta; \\
    %         &1, &&\text{~otherwise~}. \\
    %     \end{aligned}
    %     \right.
    % \end{equation*}
    There exists a mechanism that achieves the above approximation ratio.
\end{corollary}

Consider an environment $\mathbf{e} = ((\varepsilon, \varepsilon), (0, \frac{1}{\varepsilon} - \varepsilon))$, where $\varepsilon \in (0, 1)$ is a small positive.
We know that $0 < M < \Delta$ in this case.
Based on Corollary \ref{cor_apx_n2}, the lower bound of the social cost approximation is at least $\frac{1}{2\varepsilon^{2}}$, which can be arbitrarily large.
Therefore, we finish the proof of Theorem~\ref{thm: lb}.

\subsection{A Class of Strategyproof and Anonymous Mechanisms for any \(m\) and \(n\)}

Although there is no hope of designing strategyproof and anonymous mechanisms with a good approximation ratio, we show in this subsection that the mechanisms in Definition \ref{def_sp_mech} can be extended to arbitrary $m$ and $n$ while preserving desirable properties.

A \emph{best assignment of a position} $x \in \mathbb{R}$ is defined as the assignment that minimizes the cost of an agent at position $x$.
In the case of ties, we assume that they are broken by an arbitrary fixed rule.
It can be seen that the best assignment of a certain position must be one in which all agents are assigned to the same facility.
Denote $x$'s best assignment by $(\tau(x), \tau(x), \dots, \tau(x))$, where $\tau(x) \in [m]$.
We also call $(\tau(x_i), \tau(x_i), \dots, \tau(x_i))$ the \emph{best assignment of agent} $i$.
% By a closer inspection of the five types of mechanisms in Definition~\ref{def_sp_mech}, we distinguish the trivial mechanism from the other four types of mechanisms by the notion of \emph{unanimity}, which is a natural and appealing assumption which merely requires an assignment which is optimal to all agents (if it exists) to be the output (see, e.g., \cite{aswal2003dictatorial,sen2001another,pramanik2015further}). 
% \begin{definition}
%     A mechanism \(f\) is unanimous if for any environment \(\mathbf{e}\), any assignment \(\mathbf{s}\) and any agent position profile \(\mathbf{x}\) such that each agent's best assignment is \(\mathbf{s}\), then \(f(\mathbf{x},\mathbf{e})=\mathbf{s}\).
% \end{definition}
By a closer inspection of the five types of mechanisms in Definition~\ref{def_sp_mech}, we distinguish the trivial mechanism from the other four types of mechanisms by the notion of \emph{unanimity}: A mechanism \(f\) is \emph{unanimous} if for any environment \(\mathbf{e}\), any assignment \(\mathbf{s}\) and any agent position profile \(\mathbf{x}\) such that each agent's best assignment is \(\mathbf{s}\), then \(f(\mathbf{x},\mathbf{e})=\mathbf{s}\).

Obviously, the trivial mechanism is not unanimous.
It is easy to verify that the other mechanisms in Definition \ref{def_sp_mech} are unanimous.
Moreover, we have the following observation: Mechanisms of type II or IV always output the best assignment of the leftmost agent, and mechanisms of type II or IV always output that of the rightmost agent. We generalize such observation:

\begin{definition} [rank mechanisms] \label{def_rank_mech}
    For any $k\in [n]$, let $f^k$ be the mechanism such that for any environment $\mathbf{e} \in \mathcal{E}$ and any agent position profile $\mathbf{x} \in \mathbb{R}^n$, \(f^k(\mathbf{x},\mathbf{e})=(\tau(\theta_k),\tau(\theta_k),\ldots, \tau(\theta_k)),\)
    where $\theta_k$ is the $k$th smallest position in $\mathbf{x}$.
    We call $f^k$ the $k$-rank mechanism.
\end{definition}

% \begin{lemma}[\(\clubsuit\)] \label{lem: an_un}
%     $k$-rank mechanism is anonymous and unanimous for any $k \in [n]$.
% \end{lemma}

% \begin{lemma}[\(\clubsuit\)] \label{lem: sp}
%     $k$-rank mechanism is strategyproof for any $k \in [n]$.
% \end{lemma}

\begin{theorem}[\(\clubsuit\)] \label{thm: sp+an+un}
    $k$-rank mechanism $f^k$ is strategyproof, anonymous, and unanimous for any $k\in [n]$.
\end{theorem}

This entails that the preference domain induced by the cost functions of agents' is not governed by Gibbard-Satterthwaite impossibility \cite{gibbard1973manipulation,satterthwaite1975strategy}.
% , or precisely, not in the dictatorial domain (see \cite{aswal2003dictatorial,sen2001another,pramanik2015further} for details).

\section{Conclusions}
In this paper, we study the one-dimensional \myGame{} from a game-theoretical standpoint, where strategic agents aim to minimize their costs.
We explore two settings, each with a specific emphasis on equilibrium computation and mechanism design, respectively.
In the first setting where agents can directly select any facility, we demonstrate that in polynomial time we can find a PNE, of which the social cost is a $\ln n$-approximation to the optimal social cost.
In the second setting where agents report their positions to a mechanism for an assignment to facilities, we establish that no strategyproof and anonymous mechanism can achieve a bounded social cost approximation ratio. Moreover, we identify a class of mechanisms that are unanimous, strategyproof, and anonymous.

The most intriguing open problem is to resolve the complexity of computing a PNE in the Metric \myGame{}.
Another interesting problem is to investigate the hardness of finding the optimal PNE.
For the mechanism design setting, an interesting direction is to extend the characterization for $m=n=2$ to any $m$ and $n$. 
Furthermore, it is also intriguing to explore the preference domain shaped by agents' cost function in the context of social choice. 

\bibliographystyle{splncs04}
\bibliography{mybibfile}

\clearpage
% \pagenumbering{arabic} 

\appendix

\section{Appendix for Paper ``Facility Assignment with Fair Cost Sharing: Equilibrium and Mechanism Design''}

\subsection{Proof of Lemma \ref{lm_pne_nocross}}
\begin{proof}
    Assume \(i,j\in [m]\) and \(i\leq j\).
    Suppose for contradiction that there is a PNE such that \(s_a=i\) and \(s_b=j\) and \(\ell_i>\ell_j\). By Definition \ref{def_pne}, we have
    \begin{align}
        c(x_a,\mathbf{s})=|x_a-\ell_{i}|+\frac{b_i}{n_i(\mathbf{s})} &\le |x_a-\ell_j|+\frac{b_j}{n_j(\mathbf{s})+1} \label{eq_abr};\\
        c(x_b,\mathbf{s})=|x_b-\ell_{j}|+\frac{b_j}{n_j(\mathbf{s})} &\le |x_b-\ell_i|+\frac{b_i}{n_i(\mathbf{s})+1} \label{eq_bbr}.
    \end{align}
    We distinguish between the following \(6\) cases.

    \textbf{Case 1.} \(\ell_j \le x_a < x_b \le \ell_i\).
    We have \(|x_a-\ell_i|>|x_b-\ell_i|\) and \(|x_a-\ell_j|<|x_b-\ell_j|\). By applying the substitution in \eqref{eq_abr}, we get
    \begin{align*}
        |x_b-\ell_i|+\frac{b_i}{n_i(\mathbf{s})}<|x_b-\ell_j|+\frac{b_j}{n_j(\mathbf{s})+1},
    \end{align*}
    which contradicts \eqref{eq_bbr}.

    \textbf{Case 2.} \(\ell_j < \ell_i \le x_a < x_b\). We have \(|x_b-\ell_j|=|x_a-x_b|+|x_a-\ell_j|\). 
    %and \(|x_b-x_a|\ge |x_b-\ell_i|\).
    Then by \eqref{eq_abr} we have
    \begin{align*}
        |x_b-\ell_{j}|+\frac{b_j}{n_j(\mathbf{s})} &= |x_a-x_b|+|x_a-\ell_j|+\frac{b_j}{n_j(\mathbf{s})} \\
        &> |x_a-x_b|+|x_a-\ell_j|+\frac{b_j}{n_j(\mathbf{s})+1} \\
        &> |x_b-x_a|+|x_a-\ell_{i}|+\frac{b_i}{n_i(\mathbf{s})} \\
        &\ge |x_b-\ell_i|+\frac{b_i}{n_i(\mathbf{s})},
    \end{align*}
    which contradicts \eqref{eq_bbr}.

    \textbf{Case 3.} \(\ell_j \le x_a \le \ell_i \le x_b\). 
    We have \(|\ell_i-x_a|+|x_b-\ell_j|\ge |x_a-\ell_j|+|x_b-\ell_i|\).
    Then by adding \eqref{eq_abr} and \eqref{eq_bbr}, we get 
    \begin{align*}
        &|x_a-\ell_j|+|x_b-\ell_i|+
        \frac{b_i}{n_i(\mathbf{s})} + \frac{b_j}{n_j(\mathbf{s})}
        \le \\
        &|x_a-\ell_j|+|x_b-\ell_i|+
        \frac{b_i}{n_i(\mathbf{s})+1}+ \frac{b_j}{n_j(\mathbf{s})+1},
    \end{align*}
    which implies that 
    \begin{align*}
        \frac{b_i}{n_i(\mathbf{s})} + \frac{b_j}{n_j(\mathbf{s})}
        \le
        \frac{b_i}{n_i(\mathbf{s})+1}+ \frac{b_j}{n_j(\mathbf{s})+1}.
    \end{align*}
    This is a contradiction.

    \textbf{Case 4.} \(x_a < x_b \le \ell_j < \ell_i\). In this case, we also have \(|x_a-\ell_i|>|x_b-\ell_i|\) and \(|x_a-\ell_j|<|x_b-\ell_j|\). Thus the proof is the same as in case 1.

    \textbf{Case 5.} \(x_a \le \ell_j \le x_b \le \ell_i\). In this case, we have \(|x_a-\ell_i|=|x_b-x_a|+|x_b-\ell_i|\) and \(|x_b-x_a|\ge |x_a-\ell_j|\). This is symmetric with case 2 and the proof can be finished in a similar way.

    \textbf{Case 6.} \(x_a \le \ell_j \le \ell_i \le x_b\). In this case, we also have \(|x_a-\ell_j|\le |x_a-\ell_i|\) and \(|x_b-\ell_i|\le |x_b-\ell_j|\). Thus the proof is the same as in case 2.
\hfill$\square$
\end{proof}

\subsection{Proof of Theorem \ref{thm_pne_poly}}
\label{app_pne_poly}
\begin{proof}
    Denote by \(\hat{\mathbf{s}}\) the strategy profile minimizing the potential function \eqref{eq_pfunc2}. We show that \(\hat{\mathbf{s}}\) can be found by using dynamic programming.

    For any \(S\subseteq [n]\) and \(j\in [m]\), define
    \begin{align*}
        % \phi(S,j)=\sum_{k=1}^{n_j(\hat{\mathbf{s}})}\frac{b_j}{k}+\sum_{i\in S}|x_i-\ell_j|.
        \phi(S,j)=\sum_{k=1}^{|S|}\frac{b_j}{k}+\sum_{i\in S}|x_i-\ell_j|.
     \end{align*}
    Then \(\phi(A_j(\hat{\mathbf{s}}),j)\) is the term of the summation in \eqref{eq_pfunc2} for \(\hat{\mathbf{s}}\).
    By Lemma \ref{lm_pne_consecutive} we know that for each \(j\in F_{\hat{\mathbf{s}}}\), \(A_j(\hat{\mathbf{s}})\) is consecutive.
    Thus, we can decompose the potential function \eqref{eq_pfunc2} of \(\hat{\mathbf{s}}\) into ``blocks'': we compute \(\phi(A_j(\hat{\mathbf{s}}),j)\) locally for each \(j\in F_{\hat{\mathbf{s}}}\) and then add them up.

    Let \([i,j]\) be the consecutive set of agents \(\{i, i+1, \ldots, j\}\).
    Let \(MinP(i,j,k)\) be the minimum potential function value of strategy profile satisfying (i) the agents in set \([j]\) are divided into at most \(k\) consecutive subsets; (ii) agents in each partition select the same facility; (iii) \(i\in [j]\) is in the \(k\)th or the rightmost partition. Thus the optimal potential function value to our problem is \(OPT_{tc}(n,n,m)\), i.e., \(\Phi(\hat{\mathbf{s}})=MinP(n,n,m)\).

    We will make use of a \(n\times n \times m\) array \(M\), whose entries are initially set to empty.
    We invoke Algorithm \ref{alg_pneDp} to compute \(MinP(n,n,m)\) and recover the PNE from the values stored in \(M\).

    \begin{algorithm}[htbp] 
        \SetAlgoLined
        \SetNoFillComment
        \caption{\(MinP(i,j,k)\)} \label{alg_pneDp}
        % \KwIn{\(\mathbf{x},\mathbf{b},\bm{\ell}\)}
        % \KwOut{}
        \If{\(i=0\) or \(j=0\) or \(k=0\)}{
        \KwRet{\(0\)}
        }
        \ElseIf{\(M(i,j,k)\) not empty}{
            \KwRet{\(M(i,j,k)\)}
        }
        \Else{
            \(M(i,j,k)\gets \min\left\{MinP(i-1,j,k),\min_{k'\in [k]}\{ MinP(i-1,i-1,k'-1)+\phi([i,j],k')\}\right\}\)\\
            \KwRet{\(M(i,j,k)\)}
        }
    \end{algorithm}

    We argue for correctness. 
    We compute all the values of \(\phi(S,j)\) are computed in advance in time \(\mathcal{O}(nm^2)\).
    Then it is easy to see that the running time of Algorithm \ref{alg_pneDp} is \(\mathcal{O}(n^2m^2)\). 
    The base case in line \(1\) of Algorithm \ref{alg_pneDp} is clear. The potential function is zero if there are no agents or no facilities. Suppose we want to compute \(MinP(i,j,k)\) and we have already computed \(MinP(i',j',k')\) where \(i' < i\) or \(j' < j\) or \(k' < k\). We distinguish between two cases depending on whether or not agent \(i-1\) shares a facility with agent \(j\).
    If in \(\hat{\mathbf{s}}\) agent \(i-1\) also shares a facility with \(j\), then \(MinP(i,j,k)=MinP(i-1,j,k)\). Otherwise we know that agents \([i,j]\) share the same facility \(k'\in [k]\) and the remaining agents \([i-1]\) select facilities from \([k'-1]\). In this case, we need to check all possibilities of \(k'\) to get the optimal value: \(MinP(i,j,k)=\min_{k'\in [k]}\{ MinP(i-1,i-1,k'-1)+\phi([i,j],k')\}\).
    Actually, we derive the following transition function:
    \begin{align*}
        &MinP(i,j,k) \\
        =&\min\Bigl\{MinP(i-1,j,k),\\
        &\:\:\:\:\:\:\:\:\:\min_{k'\in [k]}\{ MinP(i-1,i-1,k'-1)+\phi([i,j],k')\}\Bigr\}.
    \end{align*}
    \(\phi([i,j],k')\) can be computed in \(\mathcal{O}(n)\).
    It is easy to see that Algorithm~\ref{alg_pneDp} can be implemented in \(\mathcal{O}(n^2m^2)\) time by using memorization.
\hfill$\square$
\end{proof}

\subsection{Proof of Proposition \ref{prop_pne_lnn}}
\begin{proof}
    We can rewrite the social cost of \(\hat{\mathbf{s}}\) as 
    \begin{align*}
        \mathrm{SC}(\hat{\mathbf{s}})=\sum_{j\in F(\hat{\mathbf{s}})} b_j + \sum_{i=1}^{n} d(x_i,\hat{s}_i)
    \end{align*}
    Compare above equation with the potential function \eqref{eq_pfunc1}, we get \(\mathrm{SC}(\hat{\mathbf{s}}) \le \Phi(\hat{\mathbf{s}})\). 
    Let \(\mathcal{H}_k=(1+1/2+\cdots+1/k)\) be the \(k\)th harmonic number, by \eqref{eq_pfunc1} we have 
    \begin{align*}
        \Phi(\mathbf{\mathbf{s}^*})&= \sum_{j\in F(\mathbf{\mathbf{s}^*})}\sum_{k=1}^{n_j(\mathbf{s}^*)}\frac{b_j}{k} + \sum_{i=1}^{n} d(x_i,s^*_i)\\
        & \le \left(\mathcal{H}_n \sum_{j\in F(\mathbf{s}^*)} b_j\right) + \sum_{i=1}^{n} d(x_i,s^*_i)\\
        & \le \mathcal{H}_n \cdot\left(\sum_{j\in F(\mathbf{s}^*)} b_j + \sum_{i=1}^{n} d(x_i,s^*_i)\right) \\
        & \le \ln n \cdot \mathrm{SC}(\mathbf{s}^*).
    \end{align*}
    By definition of \(\hat{\mathbf{s}}\) we also have \(\Phi(\hat{\mathbf{s}})\le \Phi(\mathbf{s}^*)\), which leads to 
    \begin{align*}
        \mathrm{SC}(\hat{\mathbf{s}}) \le \Phi(\hat{\mathbf{s}}) \le \Phi(\mathbf{s}^*) \le \ln n\cdot \mathrm{SC}(\mathbf{s}^*).
    \end{align*}

\hfill$\square$
\end{proof}

\subsection{Proof of Lemma \ref{lm_sp_ppt}}

\begin{proof}
    %Suppose for contradiction that \(\ell_j>\ell_{j'}\). 
    Assume the true position of agent \(i\) is \(x_i\). Then, agent \(i\)'s cost when she reports truthfully and the other agents report \(\mathbf{x}_{-i}\) is 
    \begin{align*}
        c(x_i,f(\mathbf{x},\mathbf{e}))=d(x_i,\ell_j)+\frac{b_j}{n_j}.
    \end{align*}
    If she unilaterally deviates to \(x_i'\), her cost becomes
    \begin{align*}
        c(x_i,f(\mathbf{x}',\mathbf{e}))=d(x_i,\ell_j')+\frac{b_{j'}}{n_{j'}'}.
    \end{align*}
    Since \(f\) is strategyproof, we have 
    \begin{align*}
        c(x_i,f(\mathbf{x},\mathbf{e}))\le c(x_i,f(\mathbf{x}',\mathbf{e})).
    \end{align*}
    That is, 
    \begin{align} \label{eq_mnt_1}
        d(x_i,\ell_j)+\frac{b_j}{n_j} \le d(x_i,\ell_{j'})+\frac{b_{j'}}{n_{j'}'}.
    \end{align}

    Then assume the true position of agent \(i\) is \(x_i'\). Then, agent \(i\)'s cost when she reports truthfully and the other agents report \(\mathbf{x}_{-i}\) is 
    \begin{align*}
        c(x_i',f(\mathbf{x}',\mathbf{e}))=d(x_i',\ell_{j'})+\frac{b_{j'}}{n_{j'}'}.
    \end{align*}
    If she unilaterally deviates to \(x_i\), her cost becomes 
    \begin{align*}
        c(x_i',f(\mathbf{x},\mathbf{e}))=d(x_i',\ell_j)+\frac{b_{j}}{n_{j}}.
    \end{align*}
    Since \(f\) is strategyproof, we have \(c(x_i',f(\mathbf{x}',\mathbf{e})) \le c(x_i',f(\mathbf{x},\mathbf{e}))\).
    That is,
    \begin{align} \label{eq_mnt_2}
        d(x_i',\ell_{j'})+\frac{b_{j'}}{n_{j'}'} \le d(x_i',\ell_j)+\frac{b_{j}}{n_{j}}.
    \end{align}
    By applying \eqref{eq_mnt_1} \(+\) \eqref{eq_mnt_2}, we have
    \begin{align} \label{eq_mnt_3}
        d(x_i,\ell_j)+d(x_i',\ell_{j'}) \le d(x_i,\ell_j')+d(x_i',\ell_j)
    \end{align}

    To prove (P1), we assume that \(x_i < x_i'\) and \(\ell_{j'}<\ell_j\).
    For the sake of contradiction, we also assume that \((x_i,x_i') \cap (\ell_{j'}, \ell_j) \ne \emptyset\).
    Then, we can derive the following four cases: (1) \(x_i\le \ell_{j'}<x_i'\le \ell_j\); (2) \(x_i\le \ell_{j'}<\ell_j\le x_i'\); (3) \(\ell_{j'}\le x_i<x_i'\le\ell_j\); (4) \(\ell_{j'}\le x_i<\ell_j\le x_i'\). It is easy to verify that in each of these four cases, we have 
    \begin{align*}
        d(x_i,\ell_j)+d(x_i',\ell_{j'}) > d(x_i,\ell_j')+d(x_i',\ell_j), 
    \end{align*}
    which contradicts with \eqref{eq_mnt_3}.
    Thus, we have \((x_i,x_i') \cap (\ell_{j'}, \ell_j) = \emptyset\), which leads that either $x'_{i} \leq \ell_{j'}$ or \(\ell_{j} \leq x_{i}\) holds.

    To prove (P2), by \eqref{eq_mnt_1} and \eqref{eq_mnt_2} we have 
    \begin{align*}
        d(x_i',\ell_{j'})-d(x_i',\ell_j) \le \frac{b_j}{n_j}-\frac{b_{j'}}{n_{j'}'} \le d(x_i,\ell_{j'})-d(x_i,\ell_{j}).
    \end{align*}

    To prove (P3), we assume that \(j=j'\). Then by \eqref{eq_mnt_1} and \eqref{eq_mnt_2} we have \(n_j \le n_{j}'\) and 
    \(n_{j}' \le n_{j}\). Therefore, we get \(n_j = n_{j}'\).
\hfill$\square$
\end{proof}

\subsection{Proof of Lemma \ref{lm_sp_ppt_2a}}

\begin{proof}
    To prove (P4), we assume, without loss of generality, that \(x_1<x_2\le \ell_1\), since the case \(\ell_2\le x_1<x_2\) can be proved by symmetry.
    Suppose for contradiction that \(f_1(\mathbf{x})\ne f_2(\mathbf{x})\).
    Then we have the following two cases: 

    \textbf{Case 1.} \(f_1(\mathbf{x})=1, f_2(\mathbf{x})=2\).
    Consider agent \(1\) in \(\mathbf{x}=(x_1,x_2)\) deviates to \(x_2\), to be strategyproof, we have \(f(x_2,x_2)= 2\) since otherwise \(f(x_2,x_2)=1\) and agent \(1\) will benefit. 
    Then, by Lemma~\ref{lm_sp_ppt} (P2), we have 
    \begin{align} \label{eq_2a_1}
        b_1-\frac{b_2}{2}=\Delta.
    \end{align}
    Then consider agent \(2\) in \(\mathbf{x}=(x_1,x_2)\) deviates to \(x_1\), similarly, we have \(f(x_1,x_1)=1\).
    Again, by Lemma~\ref{lm_sp_ppt} (P2), we have
    \begin{align} \label{eq_2a_2}
        b_2-\frac{b_1}{2}=-\Delta.
    \end{align}
    By applying~\eqref{eq_2a_1} \(+\) \eqref{eq_2a_2}, we get \(b_1/2+b_2/2=0\), a contradiction. 

    \textbf{Case 2.} \(f_1(\mathbf{x})=2, f_2(\mathbf{x})=1\).
    Consider agent \(1\) in \(\mathbf{x}=(x_1,x_2)\) deviates to \(x_2\), to be strategyproof, we have \(f(x_2,x_2)= 1\) since otherwise \(f(x_2,x_2)=2\) and agent \(1\) will benefit. 
    Then, Lemma~\ref{lm_sp_ppt} (P2), we have 
    \begin{align} \label{eq_2a_3}
        b_2-\frac{b_1}{2}=-\Delta.
    \end{align}
    Then consider agent \(2\) in \(\mathbf{x}=(x_1,x_2)\) deviates to \(x_1\), similarly, we have \(f(x_1,x_1)=2\).
    Again, by Lemma~\ref{lm_sp_ppt} (P2), we have
    \begin{align} \label{eq_2a_4}
        b_1-\frac{b_2}{2}=\Delta.
    \end{align}
    By applying \eqref{eq_2a_3} \(+\) \eqref{eq_2a_4}, we get \(b_1/2+b_2/2=0\), a contradiction. 

    To prove P5, we assume that \((x_1,x_2)\cap (\ell_1,\ell_{2})\ne \emptyset\).
    For the sake of contradiction, we assume that \(f(\mathbf{x})=(2,1)\). 
    Then consider agent \(1\) in \(\mathbf{x}\) deviates to \(x_2\). By P1 in Lemma \ref{lm_sp_ppt} and anonymity of \(f\), we have \(f(x_2,x_2)=1\).
    Then, agent \(1\) benefits, a contradiction.
\hfill$\square$
\end{proof}

\subsection{Proof of Lemma \ref{lm_div_ppt}}
\begin{proof}
    Suppose there exists a profile $\hat{\mathbf{x}} = (\hat{x}_{1}, \hat{x}_{2})$ satisfying that $\hat{x}_{1} < \hat{x}_{2}$ and $f_{1}(\hat{\mathbf{x}}) \neq f_{2}(\hat{\mathbf{x}})$.
    We know that $f(\hat{\mathbf{x}}) = (1, 2)$ by Lemma~\ref{lm_sp_ppt_2a}.    
    Suppose for contradiction that \(f_{2}(x, \hat{x}_{2}) = 1\) for some \(x \le \hat{x}_2\).
    Then by Lemma \ref{lm_sp_ppt_2a}, we have \(f_{1}(x, \hat{x}_{2}) = 1\).
    Thus, we have \(f(x,\hat{x}_2)=(1,1)\) and \(f(\hat{x}_1,\hat{x}_2)=(1,2)\), which contradicts Lemma \ref{lm_sp_ppt} (P3).
    The other part of \eqref{diag} can be proved by symmetry. 
    %By \eqref{diag} we immediately have   \(f(\hat{x}_{1}, \hat{x}_{1}) = (1, 1)\text{~and~}f(\hat{x}_{2}, \hat{x}_{2}) = (2, 2)\).
\hfill$\square$
\end{proof}

\subsection{Proof of Lemma \ref{lm_mne0}}
\begin{proof}
    Suppose \(M\ne 0\).
    For a certain profile \((a,b)\) with \(a\le b\le 0\), by Lemma~\ref{lm_sp_ppt_2a} (P4), we can assume that \(f(a,b)=(j,j)\) for some \(j\in [2]\).
    Suppose for contradiction that \(f(a',b)\ne (j,j)\) for some \(a' \le 0\).
    By Lemma~\ref{lm_sp_ppt_2a} (P4), we have \(f(a',b)=(j',j')\). 
    Then, apply Lemma~\ref{lm_sp_ppt_2a} (P2) to profiles \((a,b)\) and \((a',b)\) we get \(M=0\), a contradiction. 
    Thus we have \(f(a',b)=(j,j)\) for any \(a'\le 0\). Similarly, we can prove \(f(a,b')=(j,j)\) for any \(b'\le 0\).

    For any \((x_1\le x_2 \le 0)\), by using above arguments, we have \(f(x_1,b)=(j,j)\), then \(f(x_1,x_2)=(j,j)\). This finishes the proof of part (i). By a similar way, we can prove part (ii).

    The case \(M\ne \Delta\) can be proved by symmetry.
\hfill$\square$
\end{proof}

\subsection{Proof of Lemma \ref{lm_mech_sp}}
\begin{proof}
    First, it is not hard to verify that each type of mechanism is anonymous since $f_{1}(x, x) = f_{2}(x, x)$ for any position $x \in \mathbb{R}$.

    Now, we prove the five types of mechanisms are all strategyproof.
    Notice that for any type of the mechanisms, $f_{1}(\mathbf{x}) = f_{2}(\mathbf{x})$ always holds.
    Thus, for facility $i \in [2]$, define
    \begin{equation*}
        F_{i} = \{ \mathbf{x} \in \mathbf{R}^{2} : f_{1}(\mathbf{x}) = f_{2}(\mathbf{x}) = i \}.
    \end{equation*}
    We only need to show that Lemma~\ref{lm_sp_ppt} (P2) holds on any two profiles $\mathbf{x} = (x_{1}, x_{2}) \in F_{i}$ and $\mathbf{x}' = (x'_{1}, x'_{2}) \in F_{i'}$ such that (1) $x_{1} = x'_{1}$ or $x_{2} = x'_{2}$; (2) $i \neq i'$.

    We consider three cases according to the types of mechanisms.
    
    \textbf{Case 1:} Type I.

    Type I is trivial strategyproof since $F_{1} = \mathbb{R}^{2}$ or $F_{2} = \mathbb{R}^{2}$.
    
    \textbf{Case 2:} Type II and type III.

    For type II, let $\mathbf{x} = (x_{1}, x_{2}) \in F_{1}$ and $\mathbf{x} = (x_{1}', x'_{2}) \in F_{2}$.
    It is clear that $x_{1} \neq x'_{1}$ and $x_{1} \leq \ell_{1}$.
    Hence we let $x_{2} = x'_{2}$ and check whether Lemma~\ref{lm_sp_ppt} (P2) holds on $\mathbf{x}$ and $\mathbf{x}'$.
    Since $M = 0$, we have
    \begin{equation*}
        \left\{
        \begin{aligned}
             &d(x_1',\ell_{2}) - d(x_1',\ell_1) = |x'_{1} - \ell_{2}| - |x'_{1} - \ell_{1}| \leq \Delta = \frac{b_{2}}{2} - \frac{b_{1}}{2}; \\
             &d(x_1,\ell_{2}) - d(x_1,\ell_{1}) = (\ell_{2} - x_{1}) - (\ell_{1} - x_{1}) = \Delta = \frac{b_{2}}{2} - \frac{b_{1}}{2}. \\
        \end{aligned}
        \right.
    \end{equation*} 
    and Lemma~\ref{lm_sp_ppt} (P2) holds.
    Therefore, we obtain that type II is strategyproof.
    We can also derive that type III is strategyproof in a similar way.

    \textbf{Case 3:} Type IV or type V.

    For type IV, let $\mathbf{x} = (x_{1}, x_{2}) \in F_{1}$ and $\mathbf{x} = (x_{1}', x'_{2}) \in F_{2}$.
    It is clear that $x_{1} \neq x'_{1}$ and $x_{1} \leq \ell_{1} + M \leq x_{2} \leq x'_{1}$.
    Hence we let $x_{2} = x'_{2}$ and check whether Lemma~\ref{lm_sp_ppt} (P2) holds on $\mathbf{x}$ and $\mathbf{x}'$.
    Since $0 < M < \Delta$, we have
    \begin{equation*}
        \left\{
        \begin{aligned}
             &d(x_1',\ell_{2}) - d(x_1',\ell_1) \leq \ell_{1} + \ell_{2} - 2x'_{1} \leq \frac{b_{2}}{2} - \frac{b_{1}}{2}; \\
             &d(x_1,\ell_{2}) - d(x_1,\ell_{1})  \geq 2x_{1} - \ell_{2} - \ell_{1} \geq \frac{b_{2}}{2} - \frac{b_{1}}{2}. \\
        \end{aligned}
        \right.
    \end{equation*} 
    and Lemma~\ref{lm_sp_ppt} (P2) holds.
    Therefore, we obtain that type IV is strategyproof.
    We can also derive that type V is strategyproof in a similar way.    
\hfill$\square$
\end{proof}

\subsection{Proof of Theorem \ref{thm: char}}
\begin{proof}
    For the sake of simplicity, we assume that $\ell_{1} = 0$ and $\ell_{2} = \Delta$.
    We consider three cases according to the value of $x^{*}$.
    
    \textbf{Case 1:} $-\infty \leq x^{*} \leq 0$. 
    By the definition of \(x^*\) we have \(f(c,c)=(2,2)\) for any \(c> 0\).
    By Lemma~\ref{lm_sp_ppt} (P1), for each profile $(x_{1}, x_{2})$ with \(0<x_1<x_2\), we have $f(x_{1}, x_{2}) = (2, 2)$.

    We claim that, if \(M\ne 0\), we have \(f(x_1,x_2)\ne (1,1)\) for any \(x_1 \le 0 < x_2\).
    Suppose for contradiction that there exists a profile \((a,b)\) with \((a \le 0<b)\) such that \(f(a,b)=(1,1)\). 
    We know that \(f(\varepsilon,b)=(2,2)\) holds, where \(\varepsilon\) is a small positive.
    By applying Lemma~\ref{lm_sp_ppt} (P2) to profiles \((a,b)\) and \((\varepsilon, b)\), we have 
    \begin{align*}
        0 \le \Delta + \frac{b_2}{2}-\frac{b_1}{2} \le 2\varepsilon.
    \end{align*}
    Let \(\varepsilon \rightarrow 0\), we get \(M=0\), a contradiction.

    Similarly, if \(L\ne 0\), we claim that \(f(x_1,x_2)\ne (1,2)\) for any \(x_1\le 0<x_2\).
    Suppose for contradiction that there exists a profile \((a',b')\) with \(a'\le 0 < x_2\) such that \(f(a',b')=(1,2)\).
    By applying Lemma~\ref{lm_sp_ppt} (P2) to profiles \((a',b')\) and \((\varepsilon, b)\), we have 
    \begin{align*}
        0 \le \Delta + \frac{b_2}{2}-b_1 \le 2\varepsilon.
    \end{align*}
    Let \(\varepsilon \rightarrow 0\), we get \(L=0\), a contradiction.

    Based on the above two claims, we additionally consider three subcases.

    \textbf{Subcase 1.1:} $L \neq 0$ and $M \neq 0$.
    
    By previous claims, we immediately have \(f(x_1,x_2)=(2,2)\) for any \(x_1\le 0<x_2\) in this subcase.
    Then, we proceed to prove that \(f(x_1,x_2)=(2,2)\) for any \(x_1\le x_2\le 0\). 
    Suppose for contradiction that there exists a profile \((a'',b'')\) with \(a''\le b''\le 0\) such that $f(x_{1}', x_{2}')\ne (2,2)$. 
    Then we have \(f(a'',b'')=(1,1)\) by Lemma~\ref{lm_sp_ppt_2a} (P4).
    Also, we have \(f(a'',\varepsilon)=(2,2)\).
    Applying Lemma~\ref{lm_sp_ppt} (P2) to the profiles \((a'',b'')\) and \((a'', \varepsilon)\), we have 
    \begin{align*}
        0 \le \Delta + \frac{b_2}{2} - \frac{b_1}{2} \le 2\varepsilon.
    \end{align*}
    Let \(\varepsilon \rightarrow 0\), we have \(M=0\), a contradiction. 
    
    Therefore, we conclude that \(f\) is type I and  \(f\equiv (2,2)\) in this subcase.

    \textbf{Subcase 1.2:} $L = 0$.
    
    In this subcase, we have \(M\ne 0\), and thus \(f(x_1,x_2)\ne (1,1)\) for any \(x_1\le0<x_2\).
    Since \(M\ne 0\), we can also prove that \(x^*\in \{-\infty, 0\}\).
    Indeed, suppose for the contradiction that \(-\infty < x^* < 0\).
    By the definition of \(x^*\) we can find a small positive \(\delta\) with \(x^*+\delta<0\) such that \(f(x^*+\delta,x^*+\delta)=(2,2)\) and \(f(x^*-\delta,x^*-\delta)=(1,1)\).
    This already contradicts Lemma~\ref{lm_mne0}.
    Therefore, we have \(x^*\in \{-\infty,0\}\).

    If \(x^*=0\), then there exists a profile \((c',c')\) with \(c'\le 0\) such that \(f(c',c')=(1,1)\).
    According to Lemma~\ref{lm_mne0}, we immediately derive that \(f(x_1,x_2)=(1,1)\) for any \(x_1\le x_2 \le 0\).
    We have shown that \(f(x_1,x_2) \neq (1, 1)\) for any $x_{1} < 0 < x_{2}$ since $M \neq 0$.
    We now show \(f_{2}(x_1,x_2) \neq 2\) for any \(x_1<0 <x_2 < R\) if \(x^*=0\).

    On one hand, assume for contradiction that there exists a profile \((a',b')\) with \(a'<0 <b'<R\) such that \(f(a',b')=(1,2)\).
    Since $f(a',a') = (1, 1)$, we applying Lemma~\ref{lm_sp_ppt} (P2) and obtain that
    \begin{align*}
        a' \le \Delta+b_{2}-\frac{b_1}{2} \le b'.
    \end{align*}
    which yields that $b' > R > L = 0$, a contradiction.
    
    On the other hand, assume for contradiction that there exists a profile \((a'',b'')\) with \(a''<0 < b'' < R\) such that \(f(a'',b'')=(2,2)\).
    We could obtain that \(f(a'',x_2)=(2,2)\) for any \(x_2>0\); otherwise, it holds \(f(a'',x_2) = (1,2)\), contradicting Lemma~\ref{lm_div_ppt}.
    This leads that \(f(a'',\varepsilon)=(2,2)\) and \(f(a'',-\varepsilon)=(1,1)\) hold for any positive \(\varepsilon\).
    Applying Lemma~\ref{lm_sp_ppt} (P2), we have
    \begin{align*}
        0 \le \Delta+\frac{b_2}{2}-\frac{b_1}{2} \le 2\varepsilon.
    \end{align*}
    Let \(\varepsilon \rightarrow 0\), we get \(M=0\), a contradiction.
    Therefore, we derive that \(f(x_1,x_2)\) could not be $(1, 1)$, $(1, 2)$, or $(2, 2)$, leading to a contradiction.
    
    Now, we can conclude that \(x^*=-\infty\), which implies that \(f(x,x)=(2,2)\) for any \(x\in \mathbb{R}\).
    Then, by Lemma~\ref{lm_mne0}, we have \(f(x_1,x_2)=(2,2)\) for any \(x_1\le x_2 \le 0\).
    Suppose for contradiction that \(f(a',b')=(1,2)\) for some \(a' < 0 < b'\).
    Observe that \(f(a',x_2)=(2,2)\) holds for any \(a' < x_2 < 0\).
    This contradicts Lemma~\ref{lm_div_ppt}.
    Thus, we have \(f(x_1,x_2)=(2,2)\) for any \(x_1<0<x_2\).
    Therefore, \(f\) is type I and \(f\equiv (2,2)\).

    \textbf{Subcase 1.3:} $M = 0$.
    In this subcase, we know \(L\ne 0\), and thus \(f(x_{1}, x_{2}) \ne (1,2)\) for any $x_{1} \le 0 < x_{2}$. 
    In other words, Thus \(f(x_1,x_2)\) can be \((1,1)\) or \((2,2)\) for any $x_{1} \le 0 < x_{2}$.
    
    Assume that $f(x_{1}, x_{2}) = (1, 1)$ for some $x_{1} \leq 0 < x_{2}$.
    We suppose for contradiction that $f(x_{1}, x) = (2, 2)$ for some $x \geq x_{1}$.
    Applying Lemma~\ref{lm_sp_ppt} (P2), we have
    \begin{align*}
        \frac{x_{2}}{2} + \frac{\Delta - |x_{2} - \Delta|}{2} \le M \leq \frac{x}{2} + \frac{\Delta - |x - \Delta|}{2}.
    \end{align*}
    If $0 < x_{2} < \Delta$, we have $x_{2} \leq M = 0$, a contradiction.
    If $x_{2} \geq \Delta$, we have $M \geq \Delta$, contradicting that $M = 0$.
    Therefore, we conclude that $f(x_{1}, x_{2}) = f(x_{1}, x_{1})$ for any $x_{1} \leq 0$, leading that $f$ is type II.    
    We note that type II covers type I with $f \equiv (2, 2)$.

    \textbf{Case 2:} $\Delta \leq x^{*} \leq +\infty$.
    By the definition of \(x^*\), we have \(f(c,c)=(1,1)\) for some \(\Delta \le c \le x^*\).
    Additionally, by Lemma~\ref{lm_sp_ppt} (P2), we have \(f(x_1,x_2)=(1,1)\) for any \(x_1\le x_2 \le \Delta\).
    Similarly to case 1, we also prove two useful claims.

    We first claim that, if \(M\ne \Delta\), we have \(f(x_1,x_2)\ne (2,2)\) for any \(x_1 \le 0 < x_2\).
    Suppose for contradiction that there exists a profile \((a'' , b'')\) with \(a'' \le 0 < b''\) such that \(f(a'', b'')=(2,2)\). 
    Then, we know \(f(a'', -\varepsilon)=(1,1)\).
    By applying Lemma~\ref{lm_sp_ppt} (P2) to profiles \((a'', b'')\) and \((a'', -\varepsilon)\), we have 
    \begin{align*}
        2\Delta-2\varepsilon \le \Delta + \frac{b_2}{2}-\frac{b_1}{2} \le 2\Delta.
    \end{align*}
    Let \(\varepsilon \rightarrow 0\), we get \(M = \Delta\), a contradiction.

    We next claims that, if \(R\ne \Delta\), it holds \(f(x_1,x_2)\ne (1,2)\) for any \(x_1\le 0<x_2\).
    Suppose for contradiction that there exists a profile \((a', b')\) with \(a' \le 0 < x_2\) such that \(f(a', b')=(1,2)\).
    By applying Lemma~\ref{lm_sp_ppt} (P2) to profiles \((a', b')\) and \((a,-\varepsilon)\), we have 
    \begin{align*}
        2\Delta-2\varepsilon \le \Delta + b_2 - \frac{b_1}{2} \le 2\Delta.
    \end{align*}
    Let \(\varepsilon \rightarrow 0\), we get \(R=\Delta\), a contradiction.

    We also consider the following three subcases and their characterization can be obtained in a similar way as in case 1.

    \textbf{Subcase 2.1:} \(R\ne \Delta\) and \(M\ne \Delta\). \(f\) is type I with \(f\equiv (1,1)\).

    \textbf{Subcase 2.2:} \(R=\Delta\). \(f\) is also type I with \(f\equiv (1,1)\).

    \textbf{Subcase 2.3:} \(M=\Delta\). \(f\) is type III.

    \textbf{Case 3:} $0 < x^{*} < \Delta$.
    
    By the definition of \(x^*\), for any \(x_1 < x_2 < x^*\), there exists a position $c$ with \(x_2 < c \le x^*\) such that \(f(c,c)=(1,1)\).
    By using Lemma~\ref{lm_sp_ppt} (P2), we have \(f_1(x_2,c)=1\).
    Moreover, by Lemma~\ref{lm_sp_ppt} (P3), we get \(f(x_2,c)=(1,1)\).
    Thus, we obtain \(f(x_1,x_2)=(1,1)\) for any \(x_1<x_2<x^*\).
    Similarly, we can prove that \(f(x_1,x_2)=(2,2)\) for any \(x^*<x_1<x_2\). 

    We now claim that, for any \(x_1 < x^* < x_2\), it holds that \(f(x_1, x_2)\ne (1,2)\).
    Suppose for contradiction that there exists a profile \((\hat{x}_1, \hat{x}_2)\) with \(\hat{x}_1< x^* <\hat{x}_2\) such that \(f(\hat{x}_1,\hat{x}_2)=(1,2)\).
    Let \((\hat{x}_1,x)\) be an arbitrary profile with \(x \le \Delta\) such that \(f(\hat{x}_1, x) = (1,1)\).
    The existence of such position \(x\) is guaranteed by the fact that \(f(\hat{x}_1, x_{2}) = (1,1)\) holds for any \(x_{2} \in (0,x^*)\).
    Applying Lemma~\ref{lm_sp_ppt} (P2) to profiles \((\hat{x}_1, \hat{x}_2)\) and \((\hat{x}_1, x)\), we get
    \begin{align*}
        \frac{x}{2}+\frac{\Delta-|x-\Delta|}{2} \le L \le \frac{\hat{x}_2}{2}+\frac{\Delta-|\hat{x}_2-\Delta|}{2}. 
    \end{align*}
    Since \(x \le \Delta\) and \(\Delta - \hat{x}_2 \le |\hat{x}_2-\Delta|\), we get \(x \le M \le \hat{x}_2\) from above inequality.
    By the arbitrariness of \(\hat{x}_2\) and \(\hat{x}_2'\), we have
    \begin{align*}
        f(\hat{x}_1,x_2)=\begin{cases}
            (1,2), &\text{ if } x_2>R;\\
            (1,1), &\text{ if } x_2<R.
        \end{cases}
    \end{align*}
    Similarly, based on \(f(\hat{x}_1,\hat{x}_2)=(1,2)\) and the profile \((x,\hat{x}_2)\) with \(x \le \Delta\) such that \(f(x, \hat{x}_2) = (2,2)\) , we can also get
    \begin{align*}
        f(x_1,\hat{x}_2)=\begin{cases}
            (2,2), &\text{ if } x_1>L;\\
            (1,2), &\text{ if } x_1<L.
        \end{cases}
    \end{align*}
    Thus, we know that $f(x_{1}, x_{2}) = (1, 2)$ implying that $x_{1} \leq L$ and $x_{2} \geq R$.
    This leads that for every profile $(x_{1}, x_{2})$ with $x_{1} < L$ and $x_{2} < R$, it holds $f(x_{1}, x_{2}) = (1, 1)$; and for every profile $(x_{1}, x_{2})$ with $x_{1} > L$ and $x_{2} > R$, it holds $f(x_1, x_2) = (2, 2)$.

    Consider any profile $(x_{1}, x_{2})$ with $L < x_{1} < x^* < x_{2} < R$, and we know that \(f_1(x_1,x_2)=f_2(x_1,x_2)\).
    Suppose that \((\bar{x}_1,\bar{x}_2)\) is a profile with $L < \bar{x}_{1} < M < \bar{x}_{2} < R$ such that \(f(\bar{x}_1,\bar{x}_2) = (1,1)\).
    Let $(\bar{x}_1,x)$ be a profile with $x > R$, and we have $f(\bar{x}_1,x) = (2, 2)$.
    Applying Lemma~\ref{lm_sp_ppt} (P2) to profiles \((\bar{x}_1,\bar{x}_2)\) and \((\bar{x}_1,x)\) with $x > R$, we would get
    \begin{align*}
        \frac{\bar{x}_2}{2} + \frac{\Delta-|\bar{x}_2-\Delta|}{2} \le M \le \frac{x}{2}+\frac{\Delta-|x-\Delta|}{2}. 
    \end{align*}
    We could derive that \(\bar{x}_2 \le M\) from the above inequality, a contradiction.
    Similarly, we can also prove that no profile in $\{(x_1,x_2)|L < x_{1} < M < x_{2} < R\}$ such that the profile is assigned to \((2,2)\).
    There is a contradiction as we have proved \(f(x_1,x_2)\notin \{(1,1),(2,2),(1,2)\}\) for any profile \((x_1,x_2)\) with $L < x_{1} < x^* < x_{2} < R$.
    Hence, we conclude that \(f(x_1,x_2)\ne (1,2)\) for any \(x_1<x^*<x_2\).

    From now on, we know that \(f(x_1,x_2)=(1,1)\) or \((2,2)\) for \(x_1<x^*<x_2\).
    If $(x_1, x_2)$ is a profile with $x_1< x^* <x_2$ such that \(f(x_1,x_2)=(2,2)\).
    Let \((x_1, x)\) be arbitrary profile with \(x \le \Delta\) such that \(f(x_1, x)=(1,1)\).
    The existence of \(x\) is guaranteed by the fact that $f(x_{1}, x) = (1, 1)$ for any \(x \in [x_{1}, x^*)\).
    Applying Lemma~\ref{lm_sp_ppt} (P2) to profiles \((x_1,x_2)\) and \((x_1,x)\), we get
    \begin{align*}
        \frac{x}{2}+\frac{\Delta-|x-\Delta|}{2} \le M \le \frac{x_{2}}{2}+\frac{\Delta-|x_2-\Delta|}{2}. 
    \end{align*}
    Since \(x\le \Delta\) and \(\Delta-x_2\le |x_2-\Delta|\), we derive that \(x \le M \le x_2\) from above inequality.
    By the arbitrariness of \(x_2\) and \(x\),  we have
    \begin{align} \label{type45-22}
        f(x_1,x)=\begin{cases}
            (2,2), &\text{ if } x > M;\\
            (1,1), &\text{ if } x < M,
        \end{cases}
    \end{align}

    If $(x_1, x_2)$ is a profile with $x_1< x^* <x_2$ such that \(f(x_1,x_2)=(1,1)\).
    We can similarly obtain that
    \begin{align} \label{type45-11}
        f(x, x_2)=\begin{cases}
            (2,2), &\text{ if } x > M;\\
            (1,1), &\text{ if } x < M,
        \end{cases}
    \end{align}

    Observe that equations \eqref{type45-22} and \eqref{type45-11} cannot hold at the same time.
    Thus, we finally get that there exists an agent $k \in [2]$, such that $f(x_{1}, x_{2}) = (k, k)$ for any profile with $x_{1} < M < x_{2}$.
    This also gives that $x^{*} = M$ in this case.

    We finally consider the value of $f(M, M)$.
    If $f(M, M) = (1, 1)$, we can derive that $f(x_{1}, M) = (1, 1)$ for any $x_{1} < M$ by Lemma~\ref{lm_sp_ppt_2a} (P4).
    In addition, we have $f(M, x_{2}) = (2, 2)$ for any $x_{2} > 0$.
    Otherwise, $f(M, x_{2}) = (1, 1)$ holds, and the agent $2$ in the profile $(M, x_{2})$ will benefit if she deviates to $M$.
    For the similar reason, if $f(M, M) = (2, 2)$, we can obtain that $f(x_{1}, M) = (1, 1)$ and $f(M, x_{2}) = (2, 2)$ for any $x_{2} \leq M \leq x_{1}$.
    Therefore, we conclude that \(f\) is type IV or V.
\hfill$\square$
\end{proof}

\subsection{Proof of Corollary \ref{cor_apx_n2}}
\begin{proof}
    For the sake of simplicity, we assume that $\ell_{1} = 0$, $\ell_{2} = \Delta$, and $b_{2} \geq b_{1}$.
    For any profile $\mathbf{x} = (x_{1}, x_{2})$ with $x_{1} \leq x_{2}$.
    Clearly, the optimal social cost should be the minimum of the three values $C_{11}(\mathbf{x})$, $C_{12}(\mathbf{x})$, and $C_{22}(\mathbf{x})$, where
    \begin{equation*}
        \left\{
        \begin{aligned}
            &C_{11}(\mathbf{x}) = b_{1} + |x_{1} - 0| + |x_{2} - 0|; \\
            &C_{12}(\mathbf{x}) = b_{1} + b_{2} + |x_{1} - 0| + |x_{2} - \Delta|; \\
            &C_{22}(\mathbf{x}) = b_{2} + |x_{1} - \Delta| + |x_{2} - \Delta|. \\
        \end{aligned}
        \right.
    \end{equation*}
    Define three zones
    \begin{equation*}
        \left\{
        \begin{aligned}
            &Z_{11} = \left\{(x_{1}, x_{2}) \in \mathbb{R}^{2} : x_{2} \leq \frac{\Delta + b_{2}}{2}\text{~and~}x_{1} + x_{2} \leq \Delta + \frac{b_{2} - b_{1}}{2} \right\}; \\
            &Z_{12} = \left\{(x_{1}, x_{2}) \in \mathbb{R}^{2} : x_{1} < \frac{\Delta - b_{1}}{2}\text{~and~}x_{2} > \frac{\Delta + b_{2}}{2} \right\}; \\
            &Z_{22} = \left\{(x_{1}, x_{2}) \in \mathbb{R}^{2} : x_{1} \geq \frac{\Delta - b_{1}}{2}\text{~and~}x_{1} + x_{2} \geq \Delta + \frac{b_{2} - b_{1}}{2} \right\}.\\
        \end{aligned}
        \right.
    \end{equation*}
    We have
    \begin{equation*}
        OPT(\mathbf{x}) = C_{ij}(\mathbf{x})~\text{~if~}\mathbf{x} \in Z_{ij}.
    \end{equation*}
    
    Now, we define
    \begin{equation*}
        \left\{
        \begin{aligned}
            &\alpha(\varepsilon) = \frac{\min\{b_1,b_2\} + \Delta + 2\varepsilon }{b_{1} + b_{2} + 2\varepsilon}; \\
            &\beta(\varepsilon) = \frac{\min\{b_1,b_2\} + \Delta + M + 2\varepsilon }{\max\{b_{1}, b_{2}\} + M + 2\varepsilon}, \\
        \end{aligned}
        \right.
    \end{equation*}
    and analyze the optimal approximation ratio over all the strategyproof and anonymous mechanisms for $m=n=2$:
    \begin{equation*}
        \gamma = \sup_{\mathbf{x} \in \mathbb{R}^{2}}\inf_{f \text{~is~one~of~types~I~to~V}}\frac{SC(\mathbf{x}, f(\mathbf{x}))}{OPT(\mathbf{x})},
    \end{equation*}
    where
    \begin{equation*}
        SC(\mathbf{x}, f(\mathbf{x})) = C_{ij}(\mathbf{x})~\text{~if~}f(\mathbf{x}) = (i, j).
    \end{equation*}
    We consider two cases according to the value of $M$.
    
    \textbf{Case 1:} $M \leq 0$ or $M \geq \Delta$.
    
    In this case, $2\Delta \geq |b_{2} - b_{1}|$.
    Thus, for any profile $\mathbf{x} \in \mathbb{R}^{2}$, it holds
    \begin{equation*}
        \begin{aligned}
            &OPT(\mathbf{x}) = C_{11}(\mathbf{x}), &&\text{~if~}M \geq \Delta; \\
            &OPT(\mathbf{x}) = C_{22}(\mathbf{x}), &&\text{~if~}M \leq 0; \\
        \end{aligned}
    \end{equation*}
    Then, one can easily find that the trivial mechanism type I achieves the optimal cost, leading to the approximation ratios being exactly $1$.
    Therefore, we obtain that $\gamma = 1$ in this case.

    \textbf{Case 2:} $0 < M < \Delta$.
    
    Notice that every strategyproof and anonymous mechanism satisfies that $f_{1}(\mathbf{x}) = f_{2}(\mathbf{x})$ for any $\mathbf{x} \in \mathbb{R}^{2}$.
    Consider the profile $(\varepsilon, \Delta - \varepsilon) \in Z_{12}$ for a small positive real $\varepsilon$.
    We can obtain that
    \begin{equation*}
        \gamma \geq \min\left\{ \frac{C_{11}(\varepsilon, \Delta - \varepsilon)}{C_{12}(\varepsilon, \Delta - \varepsilon)}, \frac{C_{22}(\varepsilon, \Delta - \varepsilon)}{C_{12}(\varepsilon, \Delta - \varepsilon)} \right\} = \alpha(\varepsilon).
    \end{equation*}
    Observe that
    \begin{equation*}
        \frac{1}{2}(\Delta + b_{2}) > M > \frac{1}{2}(\Delta - b_{1}),
    \end{equation*}
    which means that both $(M - \varepsilon, \Delta -\varepsilon)$ and $(\varepsilon, M + \varepsilon)$ belong to $Z_{12}$.
    We additionally derive that either $f(M - \varepsilon, \Delta -\varepsilon) = (1, 1)$ or $f(\varepsilon, M + \varepsilon) = (2, 2)$ holds for any strategyproof and anonymous mechanism.
    Thus, the approximation ratio satisfies that
    \begin{equation*}
        \gamma \geq \min\left\{  \frac{C_{11}(M - \varepsilon, \Delta -\varepsilon)}{C_{12}(M - \varepsilon, \Delta -\varepsilon)}, \frac{C_{22}(\varepsilon, M + \varepsilon)}{C_{12}(\varepsilon, M + \varepsilon)} \right\} = \beta(\varepsilon).
    \end{equation*}
    On the flip hand, type IV or V achieves the approximation ratio $\max\{ \alpha(0), \beta(0) \}$.
    This is because all agents are assigned to the same facility by a type IV or V mechanism and it holds
    \begin{equation*}
        \left\{
        \begin{aligned}
            &\min_{i \in [2]} \frac{C_{ii}(\mathbf{x})}{C_{11}(\mathbf{x})} \leq \frac{b_{2} + \Delta + M}{b_{1} + M}; \\
            &\min_{i \in [2]} \frac{C_{ii}(\mathbf{x})}{C_{12}(\mathbf{x})} \leq \frac{\min\{b_1,b_2\} + \Delta}{b_{1} + b_{2}}; \\
            &\min_{i \in [2]} \frac{C_{ii}(\mathbf{x})}{C_{22}(\mathbf{x})} \leq \frac{b_{1} + \Delta + M}{b_{2} + M}. \\
        \end{aligned}
        \right.
    \end{equation*}
    Therefore, we know that mechanisms of type IV or V achieves the approximation ratio $\max\{ \alpha(0), \beta(0) \}$.

    Finally, we complete the proof by setting $\varepsilon \to 0$.
\hfill$\square$
\end{proof}

\subsection{Proof of Theorem \ref{thm: sp+an+un}}
\begin{proof}
    For any profile, all agents are assigned to the same facility.
    This already yields that $f^k$ is anonymous.

    Given an environment $\mathbf{e}$, let $\mathbf{x}\in \mathbb{R}^n$ and $\mathbf{s}\in [m]^n$ be the profile and assignment such that the best assignment of all agents in $\mathbf{x}$ is $\mathbf{s}$.
    Then, for any $k \in [n]$, the best assignment of $\theta_k$ is also $\mathbf{s}$.
    Thus, by Definition~\ref{def_rank_mech}, we have $f^k(\mathbf{x},\mathbf{e})=\mathbf{s}$.
    Therefore, we derive that $f^{k}$ is unanimous.

    For a fixed environment $\mathbf{e}$, denoted by $h(x,j)$ the cost of agent at position $x$ under the assignment $f^{k} = (j, j, \ldots, j)$, i.e.,
    \begin{equation*}
        h(x,j) \coloneqq |x_{i} - \ell_{j}| + \frac{b_{j}}{n}.
    \end{equation*}
    For any facility $j \in [m]$, let $\tau^{-1}(j) \subseteq \mathbb{R}$ denote the set of positions $x$ of which the best assignment is $(j, j, \ldots,j)$.
    It is not hard to see that $\tau^{-1}(j)$ is empty or formed as a closed interval $[p_{j}, q_{j}]$ for some real numbers $p_{j}, q_{j} \in \mathbb{R}$. Consider a function
    \begin{equation*}
        H_{j,j'}(x) \coloneqq h(x, j) - h(x, j') = |x_{i} - \ell_{j}| - |x_{i} - \ell_{j'}| + \frac{b_{j} - b_{j'}}{n}.
    \end{equation*}
    Clearly, $H_{j,j'}(x)$ is non-decreasing (resp. non-increasing) if $\ell_{j} \leq \ell_{j'}$ (resp. $\ell_{j} \geq \ell_{j'}$).
    Observe that $\tau(p_{j}) = \tau(q_{j}) = j$, we have $h(p_{j'}, j') \leq h(p_{j}, j)$ and $h(p_{j}, j) \leq h(p_{j'}, j')$.
    It follows that $\ell_{j} \leq \ell_{j'}$ if and only if $p_{j} \leq p_{j'}$.
    Similarly, we can also derive that $\ell_{j} \leq \ell_{j'}$ if and only if $q_{j} \leq q_{j'}$. 
    
    In addition, we claim that if $x \leq q_{j} \leq q_{j'}$, then $h(x, j) \leq h(x, j')$.
    This is because $q_{j} \leq q_{j'}$ implying $\ell_{j} \leq \ell_{j'}$, which leads to 
    \begin{align*}
        h(x, j) - h(x, j') &= H_{j,j'}(x) \\&\leq H_{j,j'}(q_{j}) \\
        &= h(q_{j}, j) - h(q_{j'}, j') \leq 0.
    \end{align*}
    In a similar way, we can also derive that if $p_{j} \leq p_{j'} \leq x$, then $h(x, j) \geq h(x, j')$.

    Now, we are ready to prove that $f^{k}$ is strategyproof.
    Suppose for the contrary that $f^{k}$ is not strategyproof, which means that there exists an agent $i \in [n]$ that will benefit if she deviates to position $x_{i}'$.
    Assume that $\mathbf{x}'$ is the profile when agent $i$ deviates to position $x_{i}'$ and let $\theta'_{k}$ be the $k$th smallest position in $\mathbf{x}'$.
    
    For simplicity, we write $j = \tau(\theta_{k})$ and $j' = \tau(\theta'_{k})$, and thus $\theta_{k} \in \tau^{-1}(j) = [p_{j}, q_{j}]$ and $\theta'_{k} \in \tau^{-1}(j') = [p_{j'}, q_{j'}]$.
    Clearly, we have $x_{i} \neq \theta_{k}$; otherwise, $j$ is the best assignment of $i$.
    On one hand, if $x_{i} < \theta_{k}$, it holds that $x'_{i} \geq \theta_{k}$; otherwise, the assignment would not change.
    It follows that $x_{i} \leq x'_{i}$.
    Additionally, the position of other agents does not change, which leads to $\theta_{k} \leq \theta'_{k}$.
    Since $h(\theta_{k}, j) \leq h(\theta_{k}, j')$ and $h(\theta'_{k}, j') \leq h(\theta'_{k}, j)$, we know $H_{j,j'}(x)$ is non-decreasing.
    It follows that $\ell_{j} \leq \ell_{j'}$, and thus $x_{i} \leq q_{j} \leq q_{j'}$ holds.
    However, based on the above claim, we get $h(x_{i}, j) \leq h(x_{i}, j')$, contradicting that agent $i$ will benefit.
    On the other hand, if $x_{i} > \theta_{k}$, it holds that $x'_{i} \leq \theta_{k}$; otherwise, the assignment would not change.
    It follows that $x'_{i} \leq x_{i}$.
    We can obtain that $h(x_{i}, j) \leq h(x_{i}, j')$ by symmetry.
    This also leads to a contradiction and our proof is completed.
\hfill$\square$
\end{proof}

\end{document}